\documentclass{aa}        
\usepackage{graphicx} 
\usepackage{txfonts} 
\begin{document}    
   \title{Dynamics during outburst} 
   \subtitle{VLTI observations of the young eruptive star V1647\,Ori during its \\ 
   2003-2006 outburst\thanks{Based on observations made with  
   ESO telescopes at the Paranal Observatory under program IDs 274.C-5026 and 076.C-0736. 
   In addition, this work is based in part on archival data obtained with the Spitzer Space Telescope,  
   which is operated by the Jet Propulsion Laboratory, California Institute of Technology under a  
   contract with NASA.}} 
 
	\titlerunning{VLTI observations of V1647\,Ori during the 2003-2006 outburst} 
 
   \author{L. Mosoni\inst{1,2}, N. Sipos\inst{1,3}, P. \'Abrah\'am\inst{1}, A.~Mo\'or\inst{1}, 
	\'A.~K\'osp\'al\inst{4,5}\thanks{ESA fellow},
	  Th. Henning\inst{2}, A. Juh\'asz\inst{2,4}, M. Kun\inst{1}, 
	  Ch.~Leinert\inst{2}, S.P.~Quanz\inst{6}, Th.~Ratzka\inst{7}, 
	  A.A.~Schegerer\inst{2}, R.~van~Boekel\inst{2}, \and S.~Wolf\inst{2,8} 
          } 
 
	\authorrunning{L. Mosoni et al.} 
 
   \offprints{L\'aszl\'o Mosoni} 
 
%   \institute{Konkoly Observatory of the Hungarian Academy of Sciences, 
%	P.O. Box 67, H-1525 Budapest, Hungary, \email{mosoni@konkoly.hu}      
      \institute{Konkoly Thege Mikl\'os Astronomical Institute, Research Centre for Astronomy and Earth Sciences, Hungarian Academy of Sciences, 
	P.O. Box 67, H-1525 Budapest, Hungary, \email{mosoni@konkoly.hu} 
         \and 
%   	     Max-Planck-Institut f\"ur Astronomie, K\"onigstuhl 17, D-69117 Heidelberg, Germany 
   	     Max Planck Institute for Astronomy, K\"onigstuhl 17, D-69117 Heidelberg, Germany 
         \and 
	     Universit\"at G\"ottingen, Institut f\"ur Astrophysik, Friedrich-Hund-Platz 1, D-37077 G\"ottingen, Germany
	 \and 
	     Leiden Observatory, Leiden University, P.O. Box 9513, 2300 RA, Leiden, The Netherlands 
	 \and 
 	     Research and Scientific Support Department, European
               Space Agency (ESA-ESTEC, SRE-SA), PO Box 299, 2200 AG,
               Noordwijk, The Netherlands
	 \and
	     Institute for Astronomy, ETH Z\"urich, Wolfgang-Pauli-Strasse 27, 8093 Z\"urich, Switzerland 
	 \and 
%	     Astrophysical Institute of Potsdam, An der Sternwarte 16, D-14482 Potsdam, Germany 
	     Universit\"ats-Sternwarte M\"unchen, Ludwig-Maximilians-Universit\"at, Scheinerstr. 1, D-81679 M\"unchen, Germany  
%	     University Observatory Munich, Scheinerstrasse 1, D-81679 M\"unchen, Germany  
         \and 
             Christian-Albrechts-Universit\"at zu Kiel, Institut f\"ur Theoretische Physik und Astrophysik, 
%	     University of Kiel, Institute of Theoretical Physics and Astrophysics,  
	     Leibnizstrasse 15, D-24098 Kiel, Germany 
          }

   \date{Received XXX xx, 2012; accepted XXX xx, XXXX} 
 
% \abstract{}{}{}{}{}  
% 5 {} token are mandatory 
  
  \abstract 
   {It is hypothesized that low-mass young stellar objects undergo eruptive phases 
   during their early evolution. 
   These eruptions are thought to be caused by highly increased mass accretion from the 
disk onto the star, and therefore play an important role in the 
   early evolution of Sun-like stars, of their circumstellar disks (structure, dust 
composition), and in the formation of their planetary systems.  
   The outburst of V1647\,Ori between 2003 and 2006 offered a rare opportunity to investigate 
   such an accretion event.
   } 
   {
   By means of our interferometry observing campaign during this outburst, 
   supplemented by other observations, we investigate the temporal  
   evolution of the inner circumstellar structure of V1647\,Ori, 
   the region where Earth-like planets could be born. We also study 
the role of the changing extinction in the brightening of the object and separate it 
from the accretional brightening. 
   } 
   {We observed V1647\,Ori with MIDI on the VLTI at two epochs in this outburst. 
   First, during the slowly fading plateau phase (2005 March) and second, 
   just before the rapid fading of the object (2005 September), which ended the outburst.  
    We used the radiative transfer code MC3D to fit the interferometry data and 
    the spectral energy distributions 
    from five different epochs at different stages of the outburst.  
    The comparison of these models allowed us to trace structural changes in the system on AU-scales.
    We also considered qualitative alternatives for the interpretation of our data.
} 
   {We found that the disk and the envelope are similar 
to those of non-eruptive young stars and that the accretion rate varied during the outburst. 
   We also found evidence for the increase of the inner radii of the circumstellar disk and envelope at 
the beginning of the outburst. Furthermore, the change of the interferometric visibilities 
indicates structural changes in the circumstellar material.
We test a few scenarios to interpret these data. We also speculate that the changes are caused by the fading of the central source, which is not immediately followed by the fading of the outer regions.   
} 
   {We found that most of our results fit in the canonical picture of young eruptive 
stars.  
   Our study provided dynamical information from the regions of the innermost few~AU of the system: changes of the inner radii of the disk and envelope.
However, if the delay in the fading of the disk is responsible for the changes seen in the MIDI data, 
the effect should be confirmed by dynamical modeling.
}

 
   \maketitle 
% 
%________________________________________________________________ 
 
\section{Introduction}

The origin of the matter of low-mass stars is interstellar gas and dust, which fall onto the 
protostar through a circumstellar accretion disk. It is becoming increasingly evident 
that the accumulation of stellar material is an episodic process: the typically low 
accretion rate suddenly increases for limited periods, delivering a significant amount 
of mass onto the stellar surface (e.g., Evans et al.~\cite{evans}).  Many young 
stars exhibit temporal brightenings possibly caused by fluctuating accretion (Herbst 
et al.~\cite{herbst}, Sicilia-Aguilar et al.~\cite{aurora}). The eruptions 
of the FU Orionis- and EX 
Lupi-type classes of variable stars (FUors and EXors) may represent the 
most intense bursts (e.g., Hartmann \& Kenyon~\cite{hk96}, hereafter HK96, Herbig~\cite{herbig07}). 
Such outbursts are characterized by an optical brightening of 2-6 magnitude.  
FUor eruptions can last from some years to several decades,  
while EXor-outbursts are usually considered as shorter counterparts of those of FUors (i.e., some months long). 
A number of possible scenarios for the triggering mechanism of the outburst have been proposed: 
(a) interactions of binary or multiple systems where tidal forces disturb the circumstellar disk (Bonnell 
\& Bastien \cite{bb92}),  
(b) thermal instabilities in the disk (Bell et al. \cite{bell95}), 
(c) planet-disk interactions, where thermal instabilities in the disk are  
caused by the presence of a massive planet (Lodato \& Clarke \cite{lodato04}),  
(d) gravitational instabilities in the disk 
%due to the mass infall from the rotating envelope onto the disk 
(Vorobyov \& Basu \cite{vb06}, Zhu et al.~\cite{zhu09}), or
(e) interaction between the disk and the magnetic field (D'Angelo \& Spruit~\cite{dangelo}).
The difference between the physics of FUor and EXor-outbursts is unclear. 
Although currently only about 20 objects are classified as eruptive 
young stellar objects (e.g., HK96, Sandell \& Weintraub~\cite{sw01},  
\'Abrah\'am et al.~\cite{abr04}), on the basis of statistical 
arguments it is believed that most low-mass pre-main sequence 
stars undergo FUor and/or EXor phases. 
Thus understanding the physics of the outbursts will shed light on  
fundamental processes of the early evolution of Sun-like stars and 
their circumstellar disks (structure, dust composition) 
and in turn on the changes of the conditions in the disks that govern the formation of planetary systems.  
The importance of the outburst phase in the early disk evolution was 
demonstrated for instance by \'Abrah\'am et al.~(\cite{nature}), who discovered 
episodic crystallization of silicate grains on the disk surface due to the increased luminosity during the 2008 outburst of EX Lup, resulting in material that forms the building blocks of comets and planets.
 
One of the best-documented and most-studied outbursts in the history of eruptive stars  is 
that of \object{V1647 Ori} in 2003-2006 (see the references collected by Aspin \& 
Reipurth~\cite{ar09}).  
The low-mass central object ($M_\mathrm{star}=0.8$\,M$_{\odot}$, $T_\mathrm{star}=3800$~K,  
$R_\mathrm{star}=3.25$\,R$_{\odot}$\footnote{The radius  was recalculated from the post-outburst luminosity
considering a distance of $d=400$\,pc (Anthony-Twarog, \cite{anthony}).};
Aspin et al., \cite{aspin08}), deeply embedded in the 
dark cloud LDN 1630, brightened by $\sim$4.5\,mag in $I_\mathrm{C}$ during three months (Brice\~no et al.~\cite{briceno}) 
until it reached a peak brightness in early 2004. It illuminated a new 
conical reflection nebulosity, the McNeil Nebula (McNeil~\cite{mcneil}). 
Subsequently, the 
object slowly faded during about 20 months. Between September-October 2005 and early 2006 
the object rapidly faded to its quiescent brightness (Acosta-Pulido et al.~\cite{jap}, hereafter AP07).   
Variation of the near-infrared colors along the reddening path suggests that the 
brightening was partly caused by a temporal removal of an extinction slab (e.g., Reipurth \& Aspin~\cite{ra04}).  
Optical and near-infrared spectroscopic observations (e.g., Reipurth \& Aspin~\cite{ra04})  
showed characteristic spectral features attributed to accretion (Br$\gamma$ emission)  
and significant mass-loss (strong H$\alpha$ emission with a P\,Cyg profile).  
The variation of the accretion rate followed that of the optical brightness of the object, i.e., 
increased rapidly at the beginning of outburst, but decreased later on  (AP07). 
The variation of the X-ray emission, which must originate from accretion processes, 
also follows the optical light curve, which supports the accretion scenario (Teets et al.~\cite{teets}).

Despite the numerous studies (see references in Aspin \& Reipurth~\cite{ar09}), 
the circumstellar structure of V1647\,Ori, and especially its variations, 
have not been investigated in detail yet\footnote{The study of Muzerolle 
et al.~(\cite{muze05}) was confined to using simple 
accretion-disk/remnant-envelope models for data of a single epoch.}. 
With the aim of studying the circumstellar disk and envelope of V1647\,Ori and the dynamical processes during the
eruption, 
we fitted the spectral energy distribution (SED) of the source at several epochs.
In order to achieve a better characterization of V1647\,Ori, 
we initiated multi-epoch observations with the VLTI/MIDI mid-infrared interferometer.
Spatial information on high angular scales was collected during two epochs.
The first dataset from 2005 March was published in our preceding paper (\'Abrah\'am et 
al.~\cite{paper1}, hereafter Paper~I). Now we analyze these data together with the interferometric data from 2005 September, 
and with spectral energy distributions at different phases of the outburst.  
We performed detailed modeling of the circumstellar environment at each epoch to decide 
whether the observed temporal changes are related to the varying illumination of the disk by its central region or 
are caused by a change of the disk structure.  
The results can be directly compared with those on another young eruptive star, PV\,Cep (Kun et al.~\cite{kun}),
whose observed flux variations are partly explained by variable extinction in the innermost part of the system,
due to evaporation and recondensation of dust grains caused by the changing amounts of energy released during the outburst. 
Variations of the model parameters, together with their timescales, give insight into the dynamics of 
the outburst of the low-mass pre-main sequence star V1647\,Ori. 
 
In Section 2 we describe the MIDI, Spitzer, and ground-based photometric observations 
and their data reduction. In Section~3 we present the results: the interferometric data,  
optical and infrared light curves, and compilations of the SED.  
In Sect.~4 we describe the radiative transfer models and discuss the temporal variation 
of physical parameters. Section~5 contains our conclusions.

%__________________________________________________________________ 
 
\section{Observations and data reduction} 
 
\subsection{VLTI/MIDI} 
\label{dr:midi}
 
\begin{table*} 
\caption{Summary of observations of V1647\,Ori. 
%All I$_\mathrm{C}$ and JHK$_\mathrm{S}$ fluxes are colour-corrected.  
Notes give Spitzer AOR numbers and VLTI baseline parameters.
Synthetic photometry derived from the Spitzer spectroscopic data are marked with asterisks. 
The mid-infrared MIDI and Spitzer spectra are presented in Section 3.} 
\label{tab:log} 
\centering 
\begin{tabular}{l c c@{} c c l} 
\hline\hline 
Date         & Instrument  & Band/Wavelength & \hspace{3.8mm}Magnitude/flux [Jy]&  Notes  & References \\ 
\hline 
2007 Jan. 20 & IAC-80       & I$_\mathrm{J}$      & 19.24$\pm$0.06 mag\\ 
2007 Oct. 28 & IAC-80       & I$_\mathrm{J}$      & 19.26$\pm$0.10 mag \\ 
2008 Feb. 6  & RCC          & I$_\mathrm{C}$      & 19.54 $\pm$0.06 mag\\ 
\hline 
2007 Jan. 20 & TCS          & J      &14.60$\pm$0.06 mag&  \\ 
2007 Oct. 22 & TCS          & J      &14.58$\pm$0.06 mag&   \\ 
2007 Oct. 28 & TCS          & J      &14.44$\pm$0.05 mag&   \\ 
\hline 
2007 Jan. 20 & TCS          & H      &11.89$\pm$0.06 mag&  \\ 
2007 Oct. 22 & TCS          & H      &11.96$\pm$0.12 mag&   \\ 
2007 Oct. 28 & TCS          & H      &11.70$\pm$0.05 mag&   \\ 
\hline 
2007 Jan. 20 & TCS          & K$_\mathrm{S}$      &10.01$\pm$0.04 mag&  \\ 
2007 Oct. 22 & TCS          & K$_\mathrm{S}$      & 9.86$\pm$0.04 mag&   \\ 
2007 Oct. 28 & TCS          & K$_\mathrm{S}$      & 9.77$\pm$0.06 mag &   \\ 
\hline 
2004 Mar. 28 & Spitzer/IRAC & 3.6 $\mu$m & 1.883$\pm$0.058 &  9463808\\ 
2004 Oct. 27 & Spitzer/IRAC & 3.6 $\mu$m & 0.914$\pm$0.029 & 12260864\\ 
2005 Feb. 25 & Spitzer/IRAC & 3.6 $\mu$m & 0.945$\pm$0.034 & 11570176\\ 
2005 Mar. 25 & Spitzer/IRAC & 3.6 $\mu$m & 1.117$\pm$0.029 & 11576320\\ 
\hline 
2004 Mar. 28 & Spitzer/IRAC & 4.5 $\mu$m & 2.864$\pm$0.087 &  9463808\\ 
2004 Oct. 27 & Spitzer/IRAC & 4.5 $\mu$m & 1.353$\pm$0.044 & 12260864\\ 
2005 Feb. 25 & Spitzer/IRAC & 4.5 $\mu$m & 1.443$\pm$0.051 & 11570176\\ 
2005 Mar. 25 & Spitzer/IRAC & 4.5 $\mu$m & 1.667$\pm$0.042 & 11576320\\ 
\hline 
2004 Mar. 28 & Spitzer/IRAC & 5.8 $\mu$m & 3.713$\pm$0.111 &  9463808\\ 
2004 Oct. 27 & Spitzer/IRAC & 5.8 $\mu$m & 1.716$\pm$0.055 & 12260864\\ 
2005 Feb. 25 & Spitzer/IRAC & 5.8 $\mu$m & 1.839$\pm$0.062 & 11570176\\ 
2005 Mar. 25 & Spitzer/IRAC & 5.8 $\mu$m & 2.062$\pm$0.052 & 11576320\\ 
\hline 
2004 Mar. 28 & Spitzer/IRAC & 8.0 $\mu$m & 5.721$\pm$0.172 &  9463808\\ 
2004 Oct. 21 & Spitzer/IRS ($\star$)  & 8.0 $\mu$m & 3.71$\pm$0.19  & 12261120 \\ 
2004 Oct. 27 & Spitzer/IRAC & 8.0 $\mu$m & 2.745$\pm$0.090 & 12260864\\ 
2005 Feb. 25 & Spitzer/IRAC & 8.0 $\mu$m & 3.008$\pm$0.097 & 11570176\\ 
2005 Mar. 2  & VLTI/MIDI    & 8.7 $\mu$m & 3.5$\pm$0.5 & & \'Abrah\'am et al. (2006)\\ 
2005 Mar. 11 & Spitzer/IRS ($\star$)   & 8.0 $\mu$m & 2.32$\pm$0.12& 11569920 \\ 
2005 Mar. 24 & Spitzer/IRS ($\star$)   & 8.0 $\mu$m & 2.86$\pm$0.14& 12644096 \\ 
2005 Mar. 25 & Spitzer/IRAC & 8.0 $\mu$m & 3.217$\pm$0.083 & 11576320\\ 
2005 Sep. 19 & VLTI/MIDI    & 8.7 $\mu$m & 2.0$\pm$0.5 \\ 
\hline 
2004 Oct. 21 & Spitzer/IRS  & 4 $-$ 34 $\mu$m& & 12261120 & Quanz et al. (2007)\\ 
2005 Mar. 2  & VLTI/MIDI    & 8 $-$ 13 $\mu$m& & 56\,m / 111\degr & \'Abrah\'am et al. (2006) \\ 
2005 Mar. 11 & Spitzer/IRS  & 4 $-$ 34 $\mu$m& & 11569920 & Quanz et al. (2007)\\ 
2005 Mar. 24 & Spitzer/IRS  & 4 $-$ 34 $\mu$m& & 12644096 & Quanz et al. (2007)\\ 
2005 Sep. 19 & VLTI/MIDI     & 8 $-$ 13 $\mu$m& & 62\,m / 108\degr  \\ 
\hline 
2004 Mar. 15 & Spitzer/MIPS & 24 $\mu$m & 18.2$\pm$2.8 & 4320256  & Muzerolle et al. (2004) \\ 
2004 Oct. 14 & Spitzer/MIPS & 24 $\mu$m & 16.5$\pm$2.5 & 12260352 \\ 
2004 Oct. 21 & Spitzer/IRS ($\star$) & 24 $\mu$m & 13.68$\pm$0.71  & 12261120 \\ 
2005 Mar. 11 & Spitzer/IRS ($\star$) & 24 $\mu$m &  8.65$\pm$0.45  & 11569920 \\ 
2005 Mar. 24 & Spitzer/IRS ($\star$) & 24 $\mu$m &  9.95$\pm$0.52  & 12644096 \\ 
\hline 
2004 Mar. 15 & Spitzer/MIPS & 70 $\mu$m & 25.7$\pm$1.7 & 4320256  & Muzerolle et al. (2004)\\ 
2004 Oct. 14 & Spitzer/MIPS & 70 $\mu$m & 30.7$\pm$1.9 & 12260352 \\ 
2004 Oct. 15 & Spitzer/MIPS ($\star$)& 70 $\mu$m & 36.6$\pm$3.7 & 12260608 \\  
2005 Mar. 1  & Spitzer/MIPS ($\star$)& 70 $\mu$m & 30.6$\pm$3.1 & 11570432 \\ 
2005 Mar. 4  & Spitzer/MIPS ($\star$)& 70 $\mu$m & 32.6$\pm$3.3 & 11576576 \\ 
2006 Nov. 8  & Spitzer/MIPS & 70 $\mu$m &  2.7$\pm$0.2 & 17455360 \\ 
\hline 
2004 Oct. 15 & Spitzer/MIPS & 55 $-$ 95 $\mu$m& & 12260608 \\  
2005 Mar. 1  & Spitzer/MIPS & 55 $-$ 95 $\mu$m& & 11570432 \\ 
2005 Mar. 4  & Spitzer/MIPS & 55 $-$ 95 $\mu$m& & 11576576 \\ 
\hline 
\end{tabular} 
\end{table*}

During the 2003-2006 outburst, V1647\,Ori was successfully observed 
twice with MIDI on the VLTI (Leinert et al.~\cite{leinert03}): 
on 2005 March 2 and 2005 September 19. The projected baseline 
lengths were 56 and 62\,m with position angles of 111 and 
108$^{\circ}$. These baselines allow one to marginally resolve structures of objects with sizes of the 
mid-infrared-emitting regions of $\ge3$\,mas, which corresponds to $\ge1.2$\,AU at the distance of V1647\,Ori. 
The two observing runs were carried out on the 
UT3-UT4 baseline of the VLTI. V1647\,Ori was observed together with the spectrophotometric calibrator HD\,37160. Owing to 
the optical faintness of the object and the lack of an adequate guide star, 
MACAO (the adaptive optics system of the VLTI for the Unit Telescopes) could not support the observations. 
At the time of our first observations, the seeing was $\approx1\arcsec$. During the observations in September, 
it was better, $0.6-0.7\arcsec$. In this latter case, the atmospheric effects on the beam-overlap were less severe. 
 
The obtained sets of data consist of acquisition images with the N8.7 
filter, $8-13 \mu $m low-resolution (R=30) spectra, and interferometric measurements. 
In the data reduction we followed the general processing scheme as 
described in previous papers (e.g., Leinert et al.~\cite{leinert04} or Ratzka et al.~\cite{ratzka07}).   
MIDI data can be reduced in two independent 
ways: with the MIDI Interactive Analysis (MIA) package, which uses the 
power spectrum method, and the Expert Work Station (EWS) package, which is 
based on a coherent, linear averaging method (Chesneau~\cite{chesneau})\footnote{The EWS+MIA package can be obtained from \texttt{  
http://www.strw.leidenuniv.nl/$\sim$nevec/MIDI/}}.  
Since V1647\,Ori is not a bright object, we set up the observations in such a way that 
the scanning optical path delay (OPD) passes through the true zero OPD.  
In this case MIA should work more reliably than EWS.  
Nevertheless, we reduced the data with the 1.6 version of both softwares and found that  
the results of the two agree well. 
We also tested the results by using different mask widths, but found only a few percent effect on the visibilities.
Although we investigated all calibrator data taken on the nights of our MIDI observations, one should probably not consider those where MACAO was used
to estimate the errors of the instrumental visibilities.
Note that MACAO was not used for the observations of HD\,37160. 
Nonetheless, the results for HD\,37160 agree well with the other instrumental visibilities (within $1\sigma$) obtained the same night.
The beam overlap was good at both epochs, even despite the lack of MACAO support.
We found that the errors of the correlated fluxes are small (a few
percent), but those of the total spectra are higher ($10-15\%$) and can change
slightly with data reduction parameters. For the sake of simplicity, we applied a
$15\%$ error on the spectra, independent of wavelengths. The errors of $15\%$ on
visibility data (see Fig.~\ref{fig:sept_v}) are consistent with other
observations (c.f. Chesneau~\cite{chesneau}).

Similarly to Paper~I, we derived 8.7$\mu$m photometry of V1647\,Ori from the acquisition images, 
using the acquisition images of HD37160 for calibration (see Table~\ref{tab:log}).

In Paper~I we presented the first dataset from 2005 March reduced with MIA~1.3.  
For the sake of consistency, we re-reduced the dataset with MIA~1.6.  
We discuss the differences between the results produced by the different 
versions of MIA in Appendix~\ref{app:dr}.

\subsection{Optical and near-infrared photometry} 
 
In addition to the already existing photometric data (AP07), new near-infrared 
(NIR) J-, H-, and K$_\mathrm{S}$-band observations were carried out 
in 2007 and 2008 (Table~\ref{tab:log}). 
These data can be used to characterize the post-outburst quiescent phase object. 
The data were obtained using CAIN-2 
installed on the 1.52\,m Carlos Sanchez Telescope (TCS) at 
the Teide Observatory (Tenerife, Canary Islands, Spain). We 
also observed V1647\,Ori in the I$_\mathrm{J}$ band with the 82\,cm  
IAC-80 telescope at the Teide Observatory and in the I$_\mathrm{C}$ band with the 1\,m~RCC  
telescope at the Piszk\'estet\H o station of the Konkoly Observatory. 
The technique of observation, data reduction, and photometric 
calibration of these data were identical with our previous 
TCS, IAC-80, and RCC observations of V1647\,Ori and are described 
in detail in AP07. 
 
\subsection{Spitzer archival data} 
 
V1647 Ori was observed with the Infrared Array Camera (IRAC) 
onboard the Spitzer Space Telescope 
at 3.6, 4.5, 5.8, and 8.0\,$\mu$m at four different epochs (Table~\ref{tab:log})  
using the subarray readout mode. Offering exposure times as short as 0.02\,s, the subarray mode enables the observation of bright sources that would saturate the detector in other readout modes. IRAC data can be downloaded from the Spitzer Heritage Archive.   
We considered all IRAC \texttt{sub\underline{\,\,}array} images that provided more
reliable results than the mapping data. These were obtained on 2004 March 7
and the results were published by Muzerolle et al.~(\cite{muze05}). 
The authors derived somewhat ($\approx10\%$) lower fluxes at each observing wavelength.
All these observations except for those carried out in 2004 October consist of nine different dither positions with  
64 images at each dither position (with size of 32$\times$32 pixel). They  
were obtained with an exposure time of 0.1\,s per frame. Four dither positions were used in 2004 October. 
The data processing started with the Spitzer Science Center (SSC) basic calibrated data (BCD) produced by  
the pipeline version S18.7.  
The BCD image cubes of the 64 frames were combined into 
two-dimensional images using the irac-subcube-collapse IDL routine provided by the SSC. 
Aperture photometry was performed on 
the final images at each wavelength using a modified version of the IDLPHOT routines. 
The aperture radius was set to 3 pixels (pixel scale $1\farcs2$),  
the sky background was computed in an annulus with an inner radius of 3 pixels and a width of 4 pixels.  
In the course of the sky estimates we used an iterative sigma-clipping method, where the clipping threshold was set  
to 3$\sigma$. Following the outline of Hora et al.~(\cite{hora}), we also 
applied an array-dependent photometric correction  
and a pixel-phase correction to the measured flux densities. 
An aperture correction was then performed using the values published in the IRAC instrument 
handbook\footnote{Version~1.0, \texttt{http://ssc.spitzer.caltech.edu/irac/ iracinstrumenthandbook/}} (IIH). 
The flux 
density values measured at the different dither positions (at each band) were averaged to obtain the final photometry.  
The final uncertainties were computed by quadratically adding the measurement errors (obtained from the individual flux  
density values in each band) and an absolute calibration error of 3\% (IIH). 
No color correction was applied since it would have only a marginal effect ($<1\%$) according to our calculations.

V1647\,Ori was observed with the Infrared Spectrograph 
(IRS) of Spitzer at three different epochs. On 2004 October 21 and on 
2005 March 24 the target was measured with the  
short-low ($5.2-14.5\,\mu$m, R$\,{\sim}\,60-127$),  
short-high ($9.9-19.5\,\mu$m, R$\,{\sim}\,600$) and  
long-high ($18.7-37.2\,\mu$m, R$\,{\sim}\,600$) 
channels, while on 2005 March 11 the 
short-low and the  
long-low ($14.0-38.0\,\mu$m, R$\,{\sim}\,57-128$) channels were used.  
All these data were published by Quanz et al.~(\cite{quanz07}),  
and we considered these spectra for our modeling.  
 
V1647\,Ori was also observed with the Multiband Imaging Photometer for 
Spitzer (MIPS, Rieke et al.~\cite{mips}) in scan-map mode on 2004 March 15 and in 
photometry-mode on 2004 October 15 and 2006 November 8.  
The data analysis of the MIPS observations started with the  
pipeline (S16.1) generated  basic calibrated data (BCD) images. 
At 24\,$\mu$m we performed a flat-field correction and a background-matching  
using the MOPEX tool (Makovoz \& Marleau~\cite{mopex}). 
Following the general steps for processing MIPS 70\,$\mu$m data  
described by Gordon et al.~(\cite{gordon}), we performed column-spatial-filtering 
and  time-median-filtering on BCD images obtained at this wavelength. 
The improved BCD images were subsequently mosaiced using MOPEX.  
The output mosaic was resampled to 2\farcs45 and 4\arcsec pixel$^{-1}$ at 24 and 70\,$\mu$m.  
The MIPS $70\,\mu$m array is affected by nonlinearity at high count rates.
To correct for the 70\,$\mu$m nonlinearity effects, 
we applied the formula described by Dale et al.~(\cite{dale}, see Eq.\,1).  
At $24\,\mu$m 
V1647\,Ori is saturated, therefore we used a model PSF to fit the pixels 
that were still in the linear regime. 
At $70\,\mu$m we used a modified version of the IDLPHOT routines to obtain aperture photometry for V1647\,Ori. 
The aperture was placed at the SIMBAD position of the source  
and the background was estimated in a sky annulus between 39\arcsec and 65\arcsec.  
For images taken at the first two epochs the aperture radius was set to 18\arcsec. 
In the last image, however, the source appeared at the edge of the image, which allowed us to  
use only a smaller aperture (radius of 12\arcsec).  
The aperture-correction factors, appropriate for a 60\,K blackbody, were derived following the outline of  
Gordon et al.~(\cite{gordon}).  
The final uncertainties were computed by quadratically adding the internal error and the absolute calibration  
uncertainty (we adopted 7\%, see MIPS data 
Handbook\footnote{Version 2.0, \texttt{http://ssc.spitzer.caltech.edu/mips/ mipsinstrumenthandbook/}}).  
The final photometry and its uncertainty is listed in Table~\ref{tab:log}. 
The MIPS data in Table~\ref{tab:log} are color-corrected (by a factor of 
0.91).  
Our results for 2004 March are higher by $16\%$ and $46\%$ at 24 and 70\,$\mu$m, compared to 
those of Muzerolle et al.~(\cite{muze05}) because of the different
pipeline versions and the additional data reduction steps.

V1647\,Ori was also observed using the MIPS SED mode. Low-resolution far-infrared (55--95$\mu$m;  
$\lambda / \Delta \lambda \sim$ 15--25 )  
spectra were obtained on 2004 October 15 and on 2005 March~1~and~4. 
The data reduction of SED observations also started with the BCD images (pipeline version S16.1) and MOPEX was  
used for the processing steps (combination of data, background removal, application of the dispersion solution)
and to compile the final images with a pixel scale of 9.8{\arcsec}.  
An aperture covering 5\,pixels was used to extract the spectra from the final images. The aperture correction factors  
were taken from Lu et al.~(\cite{lu08}).  

Finally, synthetic 8.0, 24, and 70 $\mu$m photometry was derived from the IRS and 
MIPS SED data by convolving the spectra with the corresponding IRAC or MIPS filter profiles. 
The results are presented in Table~\ref{tab:log}.

\section{Results} 
 
\subsection{Optical and infrared light curves} 
\label{sec:light} 
 
\begin{figure}%[b!] 
\begin{center} 
\includegraphics[width=\columnwidth]{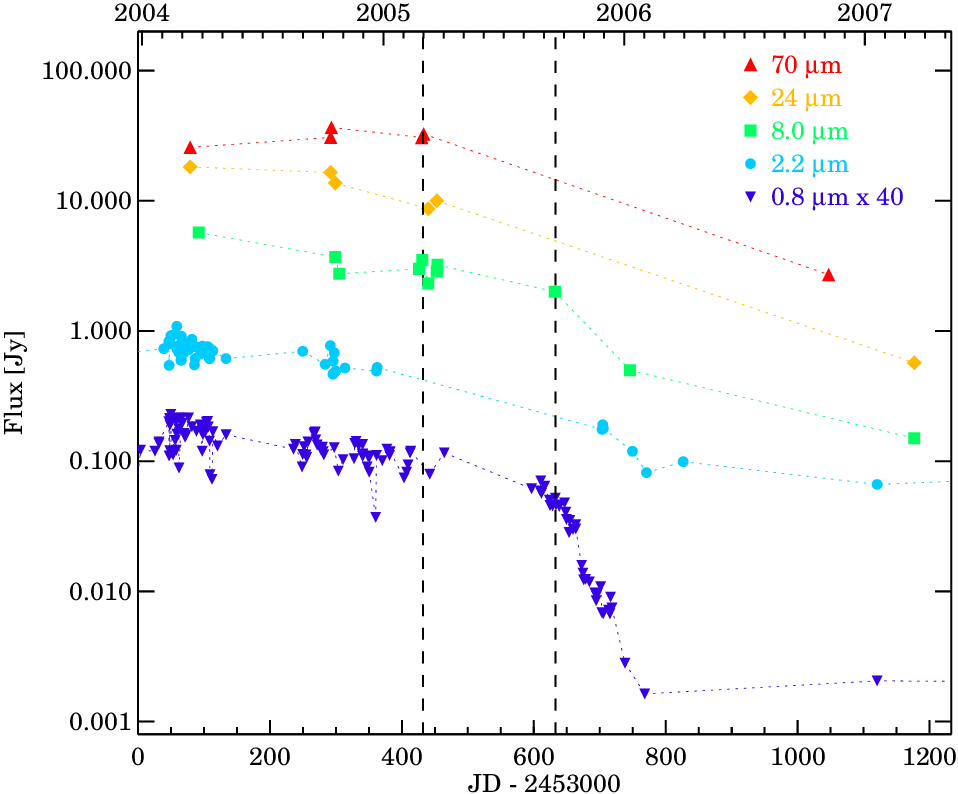} 
\caption{Light curves of V1647\,Ori. The vertical dashed lines denote the epochs of the MIDI observations. 
$I_\mathrm{C}$ (0.8$\mu$m) and K$_\mathrm{S}$ (2.2$\mu$m) band data are from AP07 and this paper.  
The I$_\mathrm{C}$ data are scaled up by a factor of 40 for the sake of better plotting.
8.0$\,\mu$m data are from Muzerolle et al.~(\cite{muze05}), Paper~I, Quanz et al.~(\cite{quanz07}), 
this paper (obtained from MIDI acqusition data),  
and Aspin et al.~(\cite{aspin08}, extrapolated from their N$^{\prime}$ band data). 
24$\,\mu$m data are from this paper and Aspin et al.~\cite{aspin08} (extrapolated from their Q$_a$ band data).  
70$\,\mu$m data are from this paper.} 
\label{img:lightcurves} 
\end{center} 
\end{figure}

Figure~\ref{img:lightcurves} shows the light curves of V1647\,Ori at 
five optical and infrared wavelengths (0.8, 2.2, 8.0, 24, and 
70$\,\mu$m) between 2004 and 2007. 
%The origin of data points are given in the figure caption.  
The overall shape of the I$_\mathrm{C}$ light curve, which is the most complete,  
can be divided into a plateau (2004 February -- 2005 September), 
a rapid fading (2005 October -- 2006 February), and the subsequent quiescent phase. 
In addition to the general slow fading during the plateau phase, short timescale low-amplitude
variations are also observed.
The rapid fading started around the epoch of the second MIDI observations. 
The available data in the K$_\mathrm{S}$ band may suggest a 
light curve similar to that of I$_\mathrm{C}$, but with a smaller amplitude. 
%The light curves at 8, 24 and 70$\,\mu$m have even less data points than the
% K$_\mathrm{S}$ one; nevertheless they may also follow a shape similar to the
% I$_\mathrm{C}$ band. 
Although the light curves at 8, 24, and 70$\,\mu$m are even more sparsely sampled, they are not inconsistent with the shape of the I$_\mathrm{C}$ band light curve.
We note that at 70\,$\mu$m,  
the object became $\sim20\%$ brighter between the 2004 March peak and 2004 October. 
During the same period, the IRAC fluxes decreased by a factor of 2.  
This suggests a time shift between the peak brightnesses at the different wavelengths.
The cause of such a shift at 70\,$\mu$m is unclear.
 
Comparing the flux values obtained in 2006 to the pre-outburst ones 
(interpolated in wavelength in Fig.~1 of \'Abrah\'am et al.~\cite{abr04}), 
we conclude that V1647\,Ori returned to quiescence after 2005.  
The 2007 fluxes at 8 and 24$\,\mu$m are even lower than 
the respective pre-outburst values.  
This could either be due to variability 
in the quiescent phase (similar as seen in the pre-outburst I$_\mathrm{C}$ data of 
Brice\~no et al.~\cite{briceno}), might be related to the beam differences 
of the different instruments, differences in the photometric systems, 
or might even indicate a different object structure after the outburst.

\subsection{Mid-infrared interferometry} 
\label{res:midi} 
 
\begin{figure*}%[ht!] 
\centerline{ 
\includegraphics[width=5.3cm]{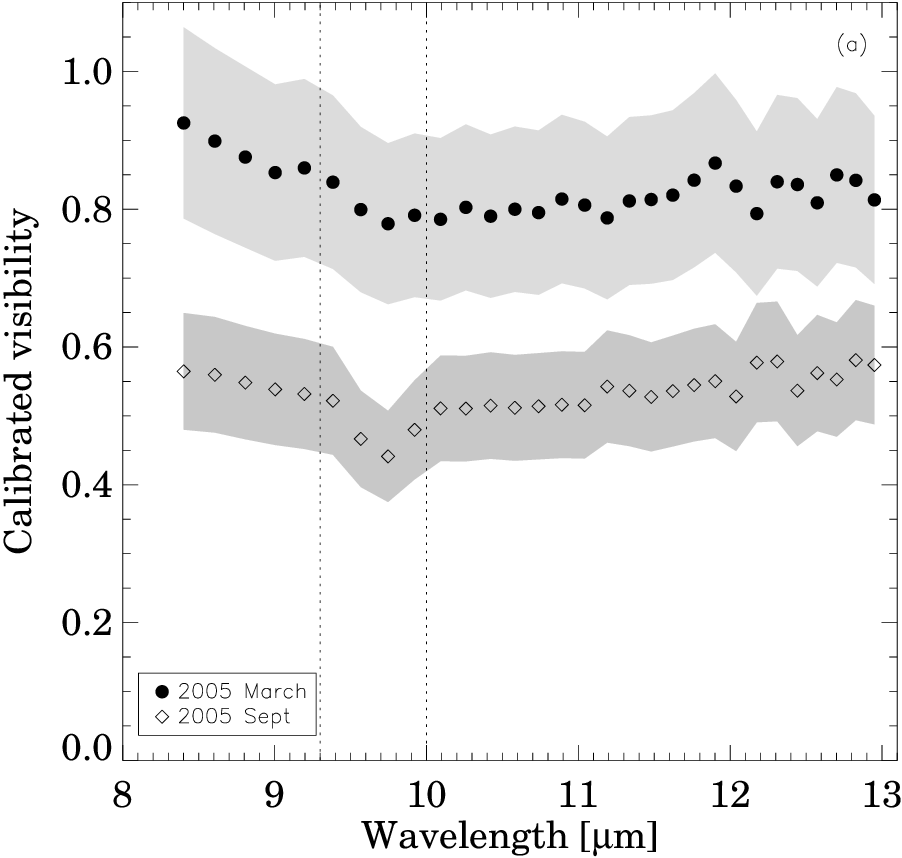} 
\includegraphics[width=5.3cm]{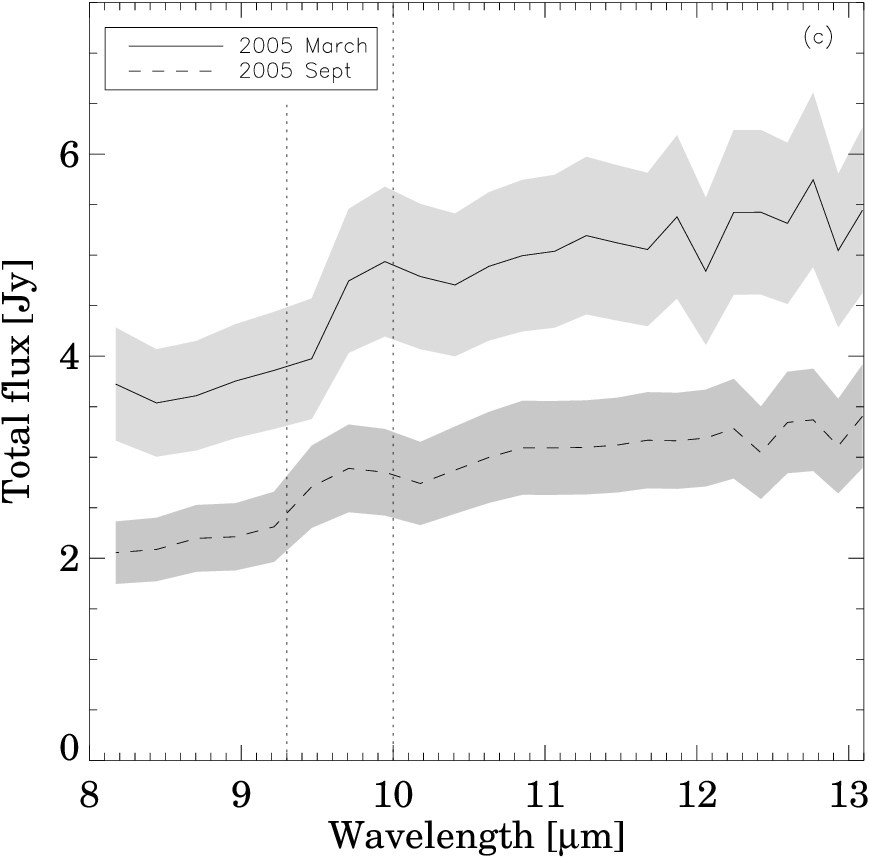} 
\includegraphics[width=5.3cm]{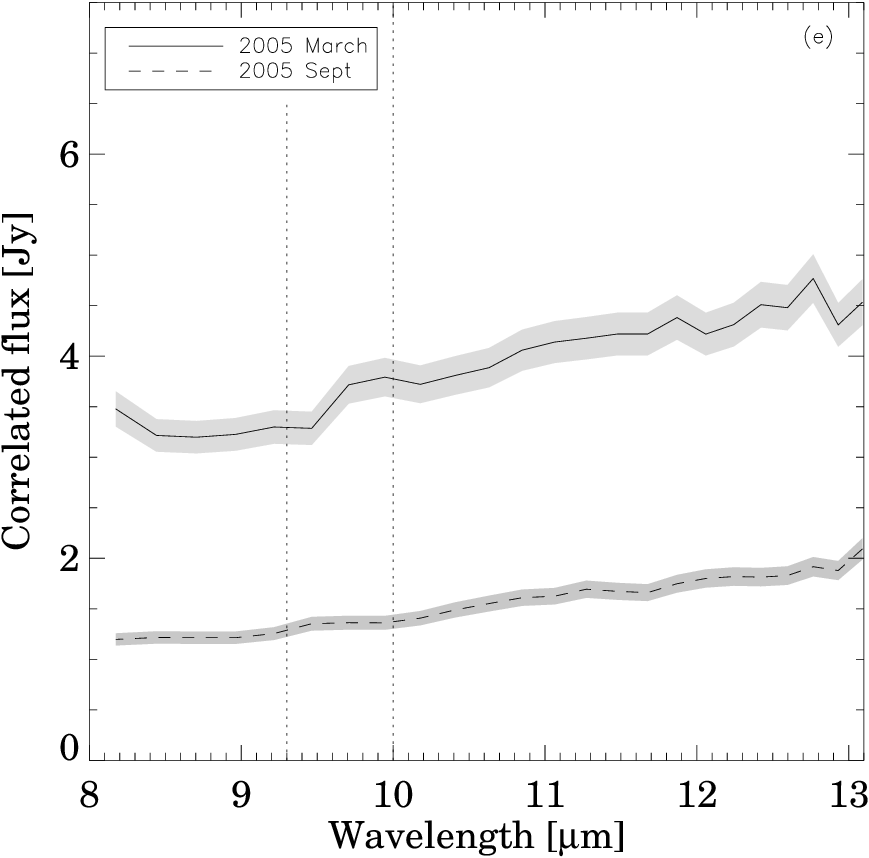}} 
\centerline{ 
\includegraphics[width=5.3cm]{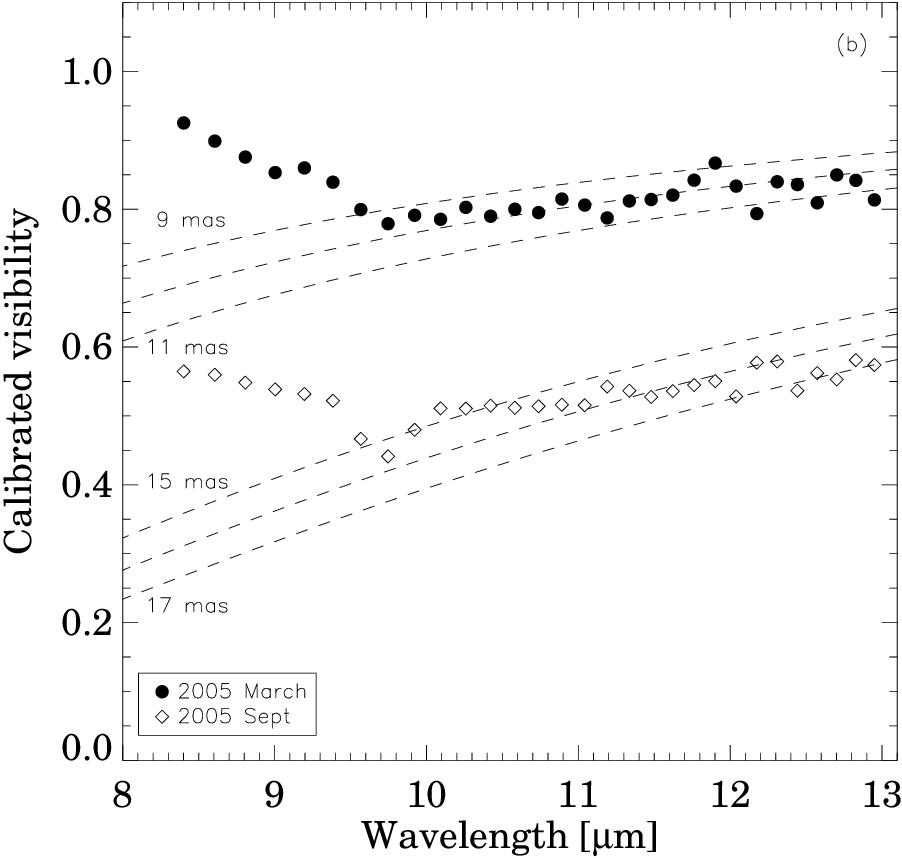} 
\includegraphics[width=5.3cm]{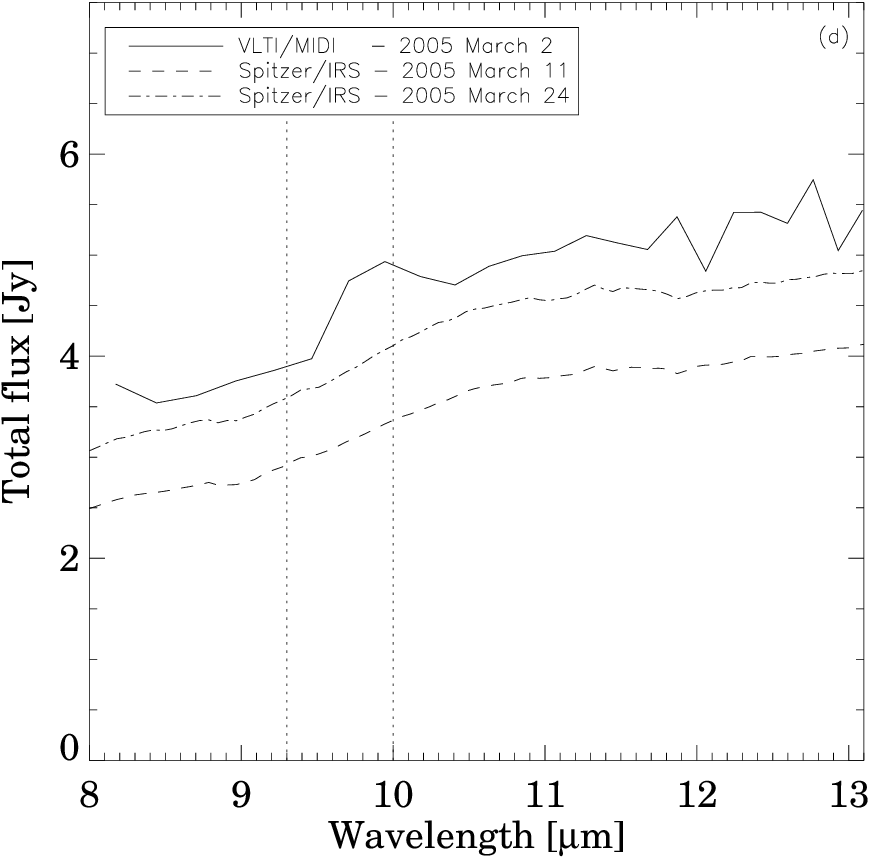} 
\includegraphics[width=5.3cm]{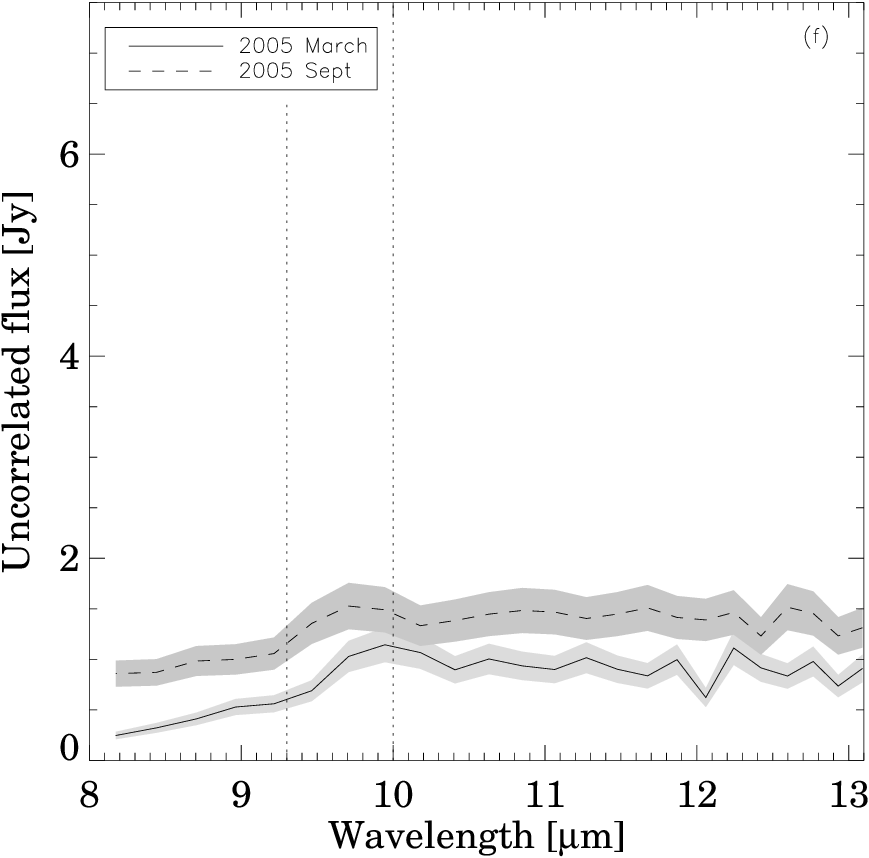}} 
\caption{Calibrated visibilities %are shown with errors of 15\% 
(a) and Gaussian size estimates (b). 
MIDI spectra are shown in panel (c). The MIDI and Spitzer/IRS spectra from 2005 March are compared in panel (d). 
The errors of the Spitzer spectra ($\sim$10\%) are not shown for the sake of clarity.  
Correlated spectra (i.e., inner disk spectra, panel (e)) and uncorrelated spectra (i.e., outer disk spectra, panel (f))  
of V1647\,Ori are also shown at two epochs. The baselines of the two observations are very similar, 
so the MIDI data of the two epochs can be compared. The shaded areas mark the errors in the panels (a,c,e,f). 
The vertical dotted lines show the ozone band that could affect the ground-based observations.} 
\label{fig:sept_v} 
\end{figure*}

Fig.~\ref{fig:sept_v} shows the calibrated visibilities obtained in 2005 March 
and September. The baselines of the two observations are 
very similar (Table~\ref{tab:log}), 
thus the two visibility curves can be directly compared.   
The difference indicates a significant change in the geometry of the mid-infrared 
emitting region of V1647\,Ori in the six months that elapsed between the two observations.  
As a first approximation, we considered Gaussian brightness distributions to estimate 
the characteristic size of the mid-infrared emitting region defined as the FWHM of 
the Gaussian. Although the object is much more resolved at the second epoch,  
and accordingly the Gaussian assumption is less adequate for size estimation,  
we converted the visibilities to the FWHM, which provided a simple comparison of the interferometric data.   
Single Gaussians with a FWHM=10 and 16\,mas, equivalent to 
4.0 and 6.4\,AU at the distance of 400\,pc, respectively, 
fit both visibility datasets well longward of $\sim10\,\mu$m.  
By fitting Gaussians to the visibilities, 
characteristic sizes of 5-7\,mas (2.0-2.8\,AU) and 11-13\,mas (4.4-5.2\,AU) can be estimated at shorter wavelengths for the two epochs.
The object was less resolved at wavelenghts shortward of $\sim10\,\mu$m at both epochs, 
probably due to an inner warm part of the system. 
Because the shape of the visibility curves is similar to those observed for other young stellar objects (YSOs) with MIDI
(e.g., Leinert et al.~\cite{leinert04}, Quanz et al.~\cite{quanz06}, Ratzka et al.~\cite{ratzka07}),
we can assume that they have similar circumstellar structures, 
which means that the radiation from the inner edge of the disk can dominate at short wavelengths.

Similarly to Paper~I, we do not see any sinusoidal variations in the spectrally resolved visibilities that are  
potentially caused by a companion. Since the baselines of the two observing runs are very similar, we can only repeat our earlier
finding that the shape of the new visibility curve suggests that no companion whose separation is between about 10 and a few hundred AU and has a brightness ratio higher than 10\% is present at the measured position angle .
 
Our two $8-13\,\mu$m MIDI spectra are shown in Fig.\ref{fig:sept_v}. 
Although the absolute level of the mid-infrared spectra decreased by a factor of 2 between the two epochs,
the shapes of the two spectra are similar.
In Fig.~\ref{fig:sept_v}~(d) we plot the first MIDI spectrum together with the two IRS spectra obtained in 2005 March. 
Variations on weekly timescales are clearly present.
Quanz et al.~(\cite{quanz07}) found weak silicate emission in the IRS spectra. 
Although the ozone band can affect the spectra at 10$\mu$m, this weak feature might also be present in the MIDI spectra.

The uncorrelated spectra are calculated as the difference of the total and the correlated ones.   
The correlated and uncorrelated spectra of an object are dominated by radiation from 
different circumstellar regions: the inner
regions of a few AU size and the outer parts (van Boekel et al.~\cite{rvb}). 
In the case of V1647\,Ori, most of the mid-infrared emission originates from the inner compact zone of the
circumstellar environment, the correlated spectrum being higher than the uncorrelated spectrum.
The correlated fluxes decreased significantly, by approximately $60\%$ between the two MIDI
observations, while the relatively low uncorrelated fluxes increased 
by about $30\%$ longward of $10\,\mu$m, and by up to $70\%$ shortward of $10\,\mu$m (Fig.~\ref{fig:sept_v}~(f)). 
These findings may suggest either a structural change in the circumstellar environment of the object or 
that the fading of the inner and the outer parts of the mid-infrared 
emitting regions of V1647\,Ori occured on different timescales. 
The correlated spectra look featureless; the weak silicate emission seems
to be associated  with the outer parts of the system.

\subsection{Spectral energy distribution} 
\label{sect:sed} 

\begin{figure*}%[th!] 
\includegraphics[width=18cm]{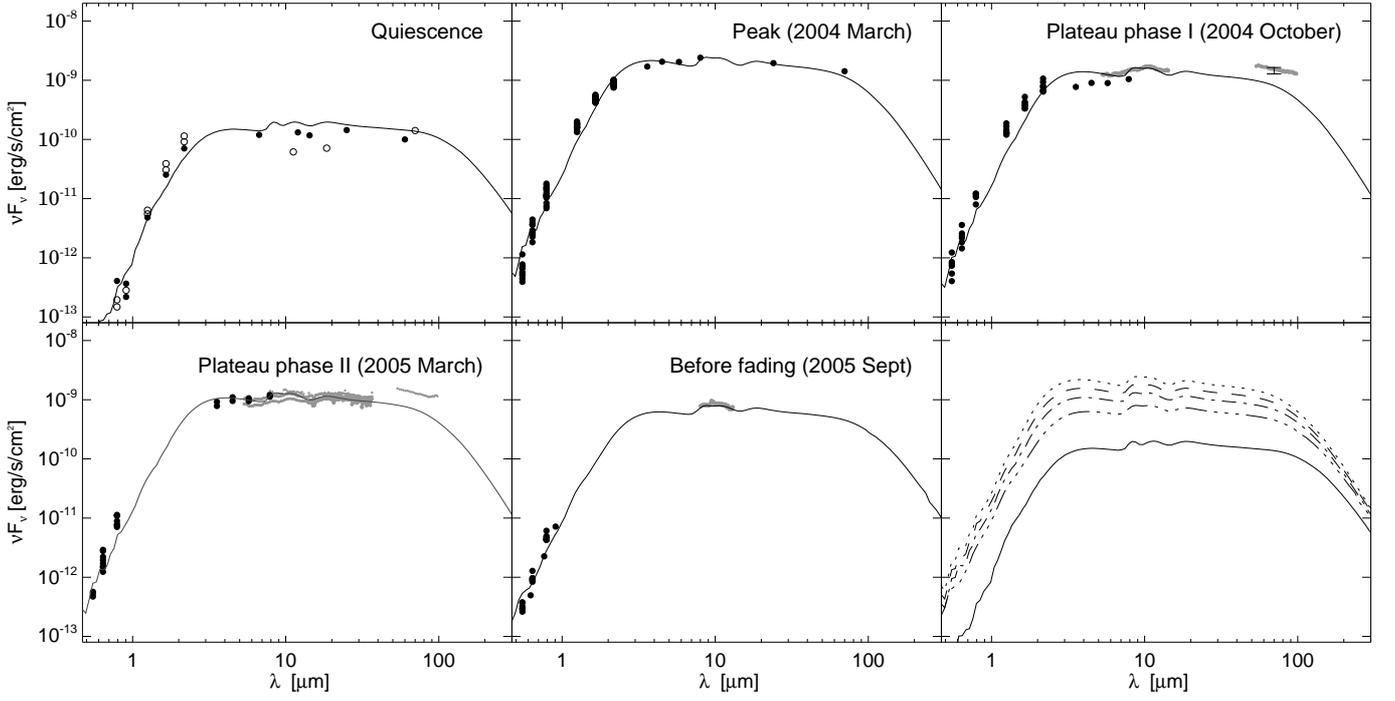} 
\caption{Spectral energy distributions of V1647\,Ori at five different epochs.  
Because of the short timescale variability at the optical and near-infared wavelengths,
we present photometry from between 2004-02-01 and 2004-04-30 (2004 March), 2004-10-01 and 2004-10-31 (2004 October),
2005-02-01 and 2005-04-30 (2005 March) and 2005-09-03 and 2005-10-03 (2005 September) in these figures.
%The best fitting 
Our models to the SEDs, where the inner radius of the dust disk is set to 0.5\,AU in quiescence and 
0.7\,AU in all outburst epochs, are overplotted with solid lines (see Sect.\ref{sec:model}). 
Between the outburst epochs only the accretion rate was changed here. Note that the model for
2005~September shown here does not fit the MIDI data (see Fig.~\ref{fig:sedvis} and 
Sect.~\ref{subsubsec:puzzle}).
The last panel shows all models. 
The decrease of the brightness of the object with time, from peak (top) to quiescence (bottom) is continuous. 
References: 
quiescence - \'Abrah\'am et al.~(\cite{abr04}), Paper~I, Aspin et al.~(\cite{aspin08}); 
peak (2004 March) - AP07, McGehee et al.~(\cite{sdss}), Muzerolle et al.~(\cite{muze05}); 
plateau phase I (2004 October) - AP07, this work, Quanz et al.~(\cite{quanz07}); 
plateau phase II (2005 March) - AP07, this work, Quanz et al.~(\cite{quanz07}); 
before the rapid fading (2005 September) - Aspin \& Reipurth~(\cite{ar09}), this work. 
} 
\label{fig:hat} 
\end{figure*}

We constructed SEDs of V1647\,Ori for several different epochs to study its circumstellar environment.  
These epochs were 2004 March (peak of the outburst), 2004 October (plateau phase 1), 
2005 March (plateau phase 2), 2005 September (shortly before the rapid fading). 
The latter two epochs are those of the MIDI observations. 
Due to the lack of simultaneous optical and near-infrared data, we included 
data points in the SEDs obtained within a period of a few weeks of the nominal date.
We also compiled a combined SED from pre- and post-outburst quiescence data (cf. Aspin et al.~\cite{aspin08}).  
Because of the sparse time coverage of observations, i.e., no simultaneous SED, we merged all available measurments 
and plotted all pre- and post-outburst data together in the quiescent SED. 
The five SEDs are plotted in Fig.~\ref{fig:hat}. 
The Spitzer IRS and MIPS SED spectra are also plotted. The absolute errors of these spectra are $\sim10$\%.
For reasons related to the radiative transfer modeling, which is described in Sect.~\ref{sec:model},  
we consider data only shortward of $100\,\mu$m. Submillimeter and millimeter data  
are used only for the estimation of the mass of the system (see Sect.\,4.1.1).

\section{Modeling} 
\label{sec:model}

In the following we fit both the SEDs and the visibilities using a radiative transfer code 
to provide a model of the circumstellar environment and the heating mechanisms. 
Our strategy for fitting the data was as follows. 
First we set up a reference model that fits the 2005 March data (Sect.~\ref{subsubsec:reference}). 
We chose this epoch to serve as a starting point because our observational data set is most complete here. 
Then we modified this initial setup as necessary to fit the data of the other epochs (Sect.~\ref{subsubsec:variation}).
The significant change of the MIDI visibilities might be explained by different scenarios, 
which are discussed separately in Sect.~\ref{subsubsec:puzzle}.

We note that first we tried to adopt the simplistic disk  
model of the V1647\,Ori system presented in our Paper~I for all epochs.  
Although the SED models were satisfactory, the model cannot reproduce 
the low visibility values measured in September 2005.

\subsection{\textbf{General overview of the model}}

\begin{figure}

\begin{center}
\includegraphics[width=\columnwidth]{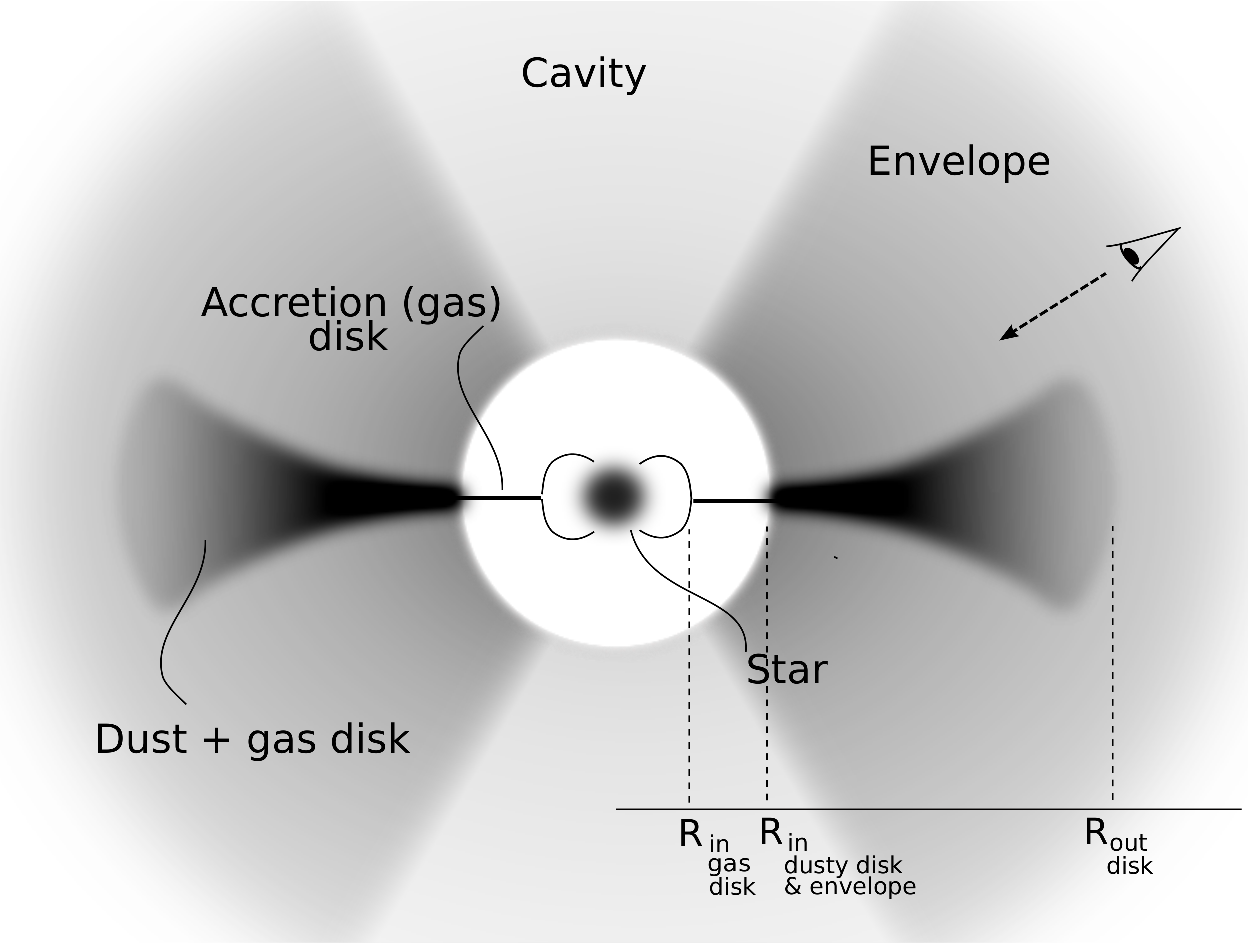}
\caption{Schematic picture of the model geometry of the object (not to scale). The line of sight crosses through the envelope. The inclination of the disk is $\vartheta\approx 60^{\circ}$, the opening angle of the conical cavity is $50^\circ$ (AP07).
}
\label{fig:rajz}
\end{center}
\end{figure}

For the radiative transfer modeling we used the Monte Carlo code MC3D (Wolf et al.~\cite{wolf99}, Wolf~\cite{wolf03}, Schegerer et al.~\cite{sch}). We assumed an axial symmetry of the circumstellar
environment, therefore we used a two-dimensional geometry in polar coordinates ($r,\theta$). 

The model system consists of a central star, an inner accretion (gas) disk, and a dusty outer disk
embedded in an envelope, from which a cone-shaped cavity is cut (Fig.~\ref{fig:rajz}).  
For the structure of the disk we assumed the density profile to be like the one used by Wood et al.~(\cite{wood}),
\[
\rho_{\mathrm{disk}} = \rho_{0,\mathrm{disk}} \left(R_*\over{\varpi}\right)^{\alpha_{\mathrm{disk}}} \exp{\left\{-\frac{1}{2}\left[ \frac{z}{h(\varpi)}\right]^2\right\}},
\]
where $\varpi$ is the radial coordinate in the disk midplane, $z$ is the vertical height and $\rho_0$ is
determined from the mass of the disk. The scale height $h(\varpi)$ increases with
radius as,
\[
h=h_0\left(\frac{\varpi}{100\,\mathrm{AU}}\right)^\beta,
\]
where $h_0$ is the scale height at a radial distance of 100\,AU.
The density profile of the envelope is spherically symmetric: 
\[
\rho_{\mathrm{env}}= \rho_{0,\mathrm{env}} \left(R_*\over{r}\right)^{\alpha_{\mathrm{env}}}.
\]
Here $r$ gives the distance from the star, and where 
$\rho_{0,\mathrm{env}}$ is defined as $f \rho_{0,\mathrm{disk}}$ at a certain arbitrary point of the disk and the envelope. 
The value of the constant $f$ has no
physical meaning, it is set so as to separately control the mass of the disk and the envelope.
The transition between these two components is assumed to be smooth.
In vertical direction the envelope starts where the density of the disk decreases to the density of the envelope.
In the cavity we used a density distribution identical to that of the envelope, but less dense by a factor of $10^{-6}$.

The temperature distribution of the system is determined by the heating sources: 
the central star, the heated dust grains, 
emitting blackbody radiation, and the accretion, consisting of an $\alpha$-type disk 
concentrated on the midplane and a hot spot on the stellar suface (for
details of accretion effects see Schegerer et al.~\cite{sch}). 
The accretion process is characterized by three parameters, the accretion rate $\dot{M}$, 
the temperature $T_{\mathrm{spot}}$ of the accreting region on the surface of the star, and the magnetic truncation radius
$R_{\mathrm{trunc}}$. 
After calculating the temperature distribution, the SED and a projected image of the system is produced at
an inclination angle $\vartheta$ with a ray-tracer. 

The circumstellar disk in our model is embedded
in a significantly larger envelope that stretches out far beyond the disk,
to 3000\,AU, in accordance with the mm-images from Tsukagoshi et al.~(\cite{tsuka}).
The inclination of the disk was fixed to be $\vartheta\approx 60^{\circ}$  
and we set the opening angle of the cavity to $50^\circ$ (AP07),
thus the line of sight to the star crosses the envelope.
Although the mid-infrared ice features indicate the presence of foreground material
(Vacca et al.~\cite{vacca}, Rettig et al.~\cite{rettig}, Quanz et al.~\cite{quanz07}), 
the ice can also be in the outer regions of the extended envelope around V1647\,Ori.
Therefore we fixed the interstellar extinction to be $A_\mathrm{V}=0$ and assumed that all extinction is circumstellar.

The 10\,$\mu$m silicate feature is too weak to determine the dust composition via model-fitting. 
Therefore we considered standard interstellar dust, a mixture of $62.5 \%$ astronomical silicate, and
$37.5 \%$ graphite (the optical parameters and grain size distribution 
were taken from Weingartner \& Draine~\cite{weingartner}). 
The shape of the continuum-subtracted features (Quanz et al.~\cite{quanz07}) 
and the presence of an envelope
support the plausibility of this assumption.
We used spherical dust grains, with a power-law size distribution 
$n(a)\propto a^{-3.5}$, for $a_{\mathrm{min}}\le a \le a_{\mathrm{max}}$, where $n(a)$ is
the number of dust particles with a radius $a$. For the minimum and maximum grain
size we used $0.005 \,\mu\mathrm{m}$ and $0.25\,\mu\mathrm{m}$. 
To limit the number of free parameters in the model fitting, 
we fixed $a_{\mathrm{min}}$ and $a_{\mathrm{max}}$. Note that increasing the maximum grain size 
to a few $\mu\mathrm{m}$ would decrease the silicate emission, 
but would not change the model SEDs significantly. 
The gas-to-dust mass ratio is assumed to be 100, while the grain mass density is set to 
$\rho_g= 2.7\,\mathrm{g cm}^{-3}$.
We have found that with the selected grain size distribution 
we underestimate the submm-to-mm flux in all our models, unless we assume a
disk mass of $\approx 1 \mathrm{M}_{\odot} $, which would result in gravitational
instabilities. In addition, the submm -- mm slope of the model is steeper than that
of the measured data, which leads to the conclusion that the system possibly
contains larger grains than the maximum size used. 
However, we think the larger grains are within the inner regions of the disk closer to the midplane
due to sedimentation, affecting the SED significantly only in the (sub-)mm regime, since this region is
optically thick at shorter wavelengths.
Consequently, we consider our model to be valid shortward of $\sim$ 100\,$\mu$m.

We estimated the mass of the system from the 1.3 mm flux (Lis et al.~\cite{lis99}), 
using the method described in Beckwith et al.~(\cite{beckwith99}). Assuming
$T=50$\,K and $\kappa_{\nu} =0.02$\,cm$^2$g$^{-1}$, we found that the total mass
of the circumstellar matter is $M_{\mathrm{tot}}=0.045$\,M$_{\odot}$. The estimate is similar to that of Andrews et al.~(\cite{andrews}). 
We considered this estimate as a fixed parameter during our model-fitting.

\subsection{\textbf{The reference model: 2005 March}}
\label{subsubsec:reference}

\begin{table*}[th!]
\caption{Parameters of the best-fitting model for 2005 March. The fitted parameters are shown in italics.}
\centering
\begin{tabular}{l  r@{\hspace{.7 mm}} ll r@{\hspace{.7 mm}} l} 
\hline\hline  
Parameters& \multicolumn{2}{c}{Final model}       & References for fixed parameters & \multicolumn{2}{c}{Ranges for fit} \\\hline  
\textbf{Stellar parameters}&&\\  
\hspace{0.5 cm}Temperature ($T_{\mathrm{star}}$) &3800            &K & Aspin et al.~(\cite{aspin08})\\  
\hspace{0.5 cm}Mass  ($M_{\mathrm{star}}$) &0.8             & $\mathrm{M}_{\odot}$ & Aspin et al.~(\cite{aspin08})\\  
\hspace{0.5 cm}Radius ($R_{\mathrm{star}}$) &3.25              &$\mathrm{R}_{\odot}$& Aspin et al.~(\cite{aspin08}), recalculated\\  
\hspace{0.5 cm}Interstellar visual extinction ($A_\mathrm{V}$)                  &0 &mag              &\\  
\textbf{Circumstellar disk parameters}       && \\  
\hspace{0.5 cm}\textit{Inner radius of dusty disk} ($R_{\mathrm{in,disk}}$) &0.7            & AU & & 0.2 - 1.5 & AU \\  
\hspace{0.5 cm}Outer radius of dusty disk ($R_{\mathrm{out,disk}}$) &500             & AU\\  
\hspace{0.5 cm}\textit{Scale height at 100 AU} ($H_0$)                          &15 & AU & & 3 - 25 & AU\\  
\hspace{0.5 cm}\textit{Flaring index} ($\beta$)                                &1.2 & & & 1 - $^9/_7$ \\  
\hspace{0.5 cm}\textit{Exponent of radial density profile} ($\alpha_{\mathrm{disk}}$) &   $-1.75$               & & & $-2.25$\,\,-\,\,$-1.5$ \\  
\hspace{0.5 cm}Total mass of disk and envelope ($M$)                   & 0.045            &$\mathrm{M}_{\odot}$ & estimated \\  
\hspace{0.5 cm}Distance ($d$)                                         & 400             &pc &  Anthony-Twarog~(\cite{anthony}), AP07 \\  
\hspace{0.5 cm}Inclination ($\vartheta$)                              & 60 &$^{\circ}$ & AP07\\  
\textbf{Circumstellar envelope parameters}&&\\  
\hspace{0.5 cm}\textit{Inner radius of dusty envelope} ($R_{\mathrm{in,env}}$) & 0.7   & AU & & 0.05 - 1.0 & AU\\  
\hspace{0.5 cm}Outer radius of dusty envelope ($R_{\mathrm{out,env}}$) &3000              &AU & Tsukagoshi et al.~(\cite{tsuka})\\  
\hspace{0.5 cm}Exponent of radial density profile ($\alpha_{\mathrm{env}}$) & $-1.5$              &\\  
\textbf{Parameters for the accretion}        &&\\  
\hspace{0.5 cm}\textit{Accretion rate} ($\dot{M}$)                    & 3.5$\times 10^{-6}$      & $\mathrm{M}_{\odot}\mathrm yr^{-1}$ & & 2 - 10$\times 10^{-6}$ & $\mathrm{M}_{\odot}\mathrm yr^{-1}$ \\  
\hspace{0.5 cm}Magnetic truncation radius ($R_{\mathrm{trunc}}$)      & 5 &$R_{\mathrm{star}}$& Calvet \& Gullbring~(\cite{calvet98})\\  
\hspace{0.5 cm}Temperature of the hot spot ($T_{\mathrm{spot}}$)                     & 6500 & K& Calvet \& Gullbring~(\cite{calvet98})\\  
\hline  
\end{tabular}  
\label{tab:para}
\end{table*}

As a first step we fitted the available photometric and interferometric data for the 2005 March epoch.
Since the SED is not composed of results of simultaneous observations and the object is variable on short
timescales (see Sect.~\ref{sect:sed}), we consider ranges of data at certain wavelengths\footnote{This holds also for the other epochs.}. 
Such a dataset prevents us from calculating correct meaningful $\chi^2$ values and comparing different models in a mathematical way.
Instead, we evaluated the plausibility of the models by visual inspection of the fits - as done in most of the cases, see e.g., Schegerer et al.~(\cite{sch}), Ratzka et al.~(\cite{ratzka09}), van Boekel et al.~(\cite{rvb10}), Juh\'asz et al.~(\cite{juhasz11}).
The model-fitting in the latter two cases is very similar to our case because the objects studied by the authors (T\,Tau and EX\,Lup, respectively) are also variable and the authors fitted merged SED datasets, as we did for the different epochs.

We set the derived model parameters against results found in the literature to filter out unphysical models that still showed good fits. The model has many parameters, and a good fraction of these cannot be determined or constrained by measurements. 
We attempted to fit the data with the extreme cases of envelope-only and disk-only models, 
but we conclude that neither of the two components can be neglected.
The parameters of the circumstellar environment are not clearly defined by the
models because the parameters of the envelope and the disk are not independent (e.g., Natta~\cite{natta}). 
To reduce the degeneracy we fixed additional system parameters and tried to consider canonical values. We set the outer radius of the disk to
500\,AU and the value of $\alpha_{\mathrm{env}}$ to $-1.5$ (Shu~\cite{shu77}). 
To reduce the number of free parameters even more, we set the inner radii of the 
disk and the envelope to be identical. 
The investigated parameter ranges are listed in Table~\ref{tab:para}. 
In the following we present our results with the fixed parameters, although our conclusions and the trends seen - based on the models we propose - are also valid for slightly different setups of the envelope-disk system. Therefore we do not present \emph{best-fit} models here, but a series of \emph{plausible} good models.

During the fitting procedure different $\dot{M}$ and $R_{\mathrm{in}}$ values were considered, with step sizes of $10^{-7}$\,M$_\odot$/yr and 0.05\,AU, respectively. The step sizes were $5-10\%$ maximum for all other epochs. 
For the reference epoch we found that
the accretion rate is $3.5\times 10^{-6}$\,M$_\odot$/yr and the inner radii  of the disk and the envelope are $R_{\mathrm{in}}=0.7$\,AU. Since
the line-of-sight to the star crosses the envelope but not the disk (Fig.~4),  the parameters of the envelope (mass, $R_{\mathrm{in}}$ and
$\alpha_{\mathrm{env}}$) strongly determine the fits in the optical-NIR regime. The thick envelope increases the extinction in the system in
order to decrease the otherwise overestimated optical-NIR model fluxes to the measured values\footnote{Note the absence of NIR photometric
data in 2005 (Figs.~\ref{img:lightcurves} and~\ref{fig:hat}).},  and it also helps reducing the height of the intrinsically strong 10\,$\mu$m
silicate emission feature. The parameters of our model are given in Table~\ref{tab:para},  the model can be seen in the upper panels
of Fig.~\ref{fig:sedvis}. The visibilities are somewhat underestimated. A better fit could be achieved, but at the cost of a poorer SED-fit.

Comparing the disk and
envelope parameters of our model with the results of similar modeling efforts of
embedded YSOs in the literature (e.g., Whitney et al.~\cite{whitney}), we may conclude that the circumstellar environment
of V1647\,Ori is rather typical, and no structural features directly linked to the eruptive
nature of the object can be identified. % ÁP %
With our model we could describe the system with a parameter setup that is plausible for young stellar objects.

\subsection{\textbf{Modeling the variations during outburst}}
\label{subsubsec:variation}

In order to fit the SEDs of the other epochs, we used the 2005 March model as an initial setup. 
At first sight, the SEDs differ only in their absolute levels, but 
their shapes are very similar (Fig.~\ref{fig:hat}). Therefore, we attempt to adjust the shift
between the levels by the variation of the accretion rate. This way we test if it is possible to keep the geometry of
the circumstellar environment identical, and only change the illumination of the system as the outburst proceeds. 
Additionally,
we tried to vary only the smallest number of parameters possible between the different epochs and restricted ourselves to
applying modifications that could be the result of a realistic process and may occur in such a short period of time.
Significant changes, affecting a large part of the disk or a complete restructuring, could happen only on longer
timescales (Chiang \& Goldreich~\cite{cg97}).
 
\begin{figure*}[htb!]
\includegraphics[width=16cm]{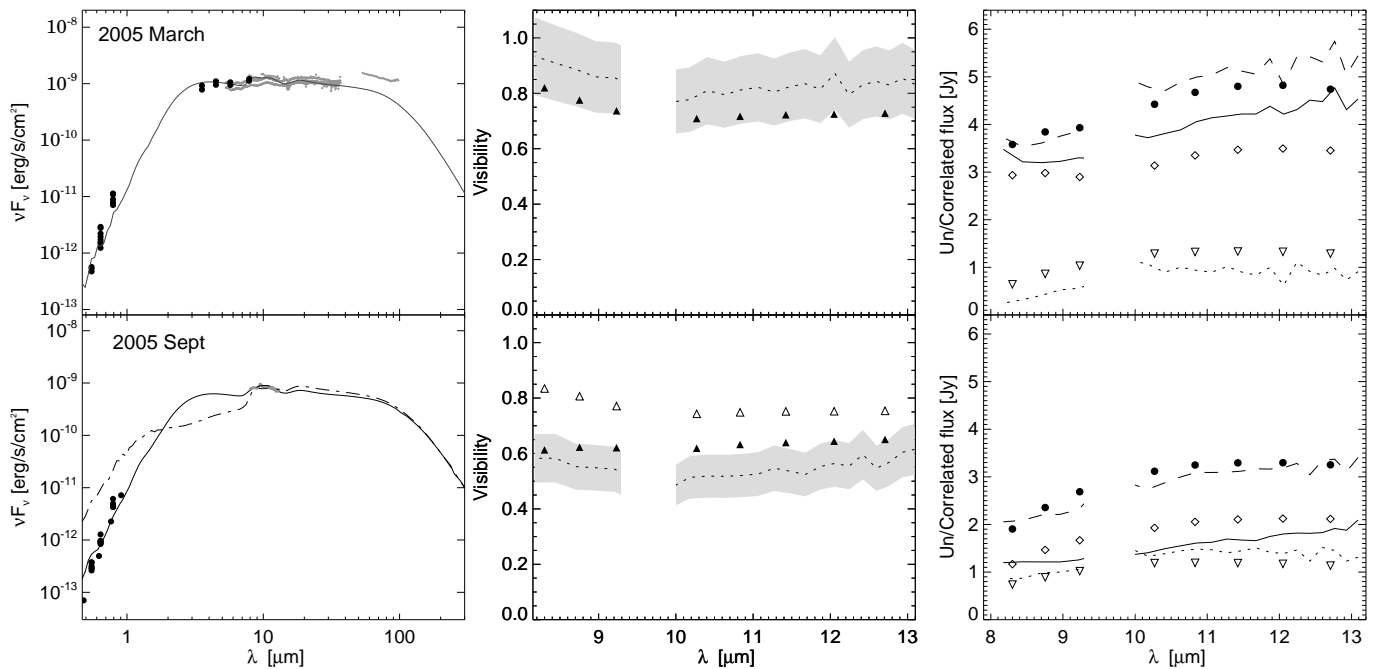}
\caption{Our models for the 2005~March and September epochs are shown in the \textit{top} and \textit{bottom row}. The SEDs and the
calibrated visibilities are shown in the \textit{left} and \textit{middle columns}. Model visibilities were calculated at
eight different wavelengths. In the \textit{bottom left panel} the solid line, in the \textit{bottom middle panel} the
open triangles show the model where only the accretion rate was decreased relative to the reference model.  This model fits the SED
quite well, but not the visibilities.  The dash-dotted line (\textit{bottom left panel}) and the filled triangles
(\textit{bottom middle panel}) show the model where the inner radius of the envelope was also increased.  As an alternative interpretation of
the mid-infrared fits,  the observed mid-infrared total, correlated and uncorrelated MIDI spectra (dashed, solid and dotted lines,
respectively) and the corresponding model values (filled circles, diamonds and downward pointing open triangles, respectively) are shown in
the \textit{right column}. 
The lower right panel refers to the case where the inner radius of the envelope increased well beyond the dust
sublimation radius, inducing larger changes in visibilities than the ones described earlier.
$R_{\mathrm{in,disk}}$ was kept unchanged.
Data probably affected by the atmospheric ozone layer are not shown.}
\label{fig:sedvis}
\end{figure*}

According to the standard picture of eruptive YSOs (HK96), 
the accretion is the main energy source in the system in outburst,
while its contribution in quiescence is less pronounced. 
The accretion rate is expected to increase from the pre-outburst phase to the peak of the outburst and 
decrease during the eruption as the source is fading. 
The modeling results confirmed this expectation. 
We found that the accretion was strongest at the peak brightness of 2004 March ($7\times10^{-6}$\,M$_{\odot}/yr$, 
see Tab.~\ref{tab:epochs}), and gradually weakened as the source approached the end of its flare-up, 
but still accreted strongly in quiescence ($3\times10^{-7}$\,M$_{\odot}/yr$). 
The fit to the mid-infrared part of the SED depends mainly on the accretion rate. 
The change found in the accretion rate is solid, since the fading can be seen clearly at these wavelengths
(see Fig.~\ref{fig:hat}).
The strategy of changing the accretion rate worked for the whole SEDs of the 
outburst epochs, but failed to reproduce  
the second MIDI interferometric data from 2005 September (see Sect.~\ref{rt:sept} 
and the middle lower panel of Fig.~\ref{fig:sedvis}).
Lowering the accretion rate to $\approx$ 5\% of the peak value reduced the fluxes considerably at all optical-IR 
wavelengths in quiescence.
This model fits the mid-infrared part of the SED well, but the excess of
the model in the optical-NIR part of the SED indicates that the
accretion rate is not the only parameter that must be adjusted.

The inner radii of the dusty components in the reference model are set to 0.7\,AU, 
which provides good fit for the other two outburst epochs in 2004.
Accordingly, the geometry of the inner part of the system was unchanged during the plateau phase.
For quiescence, 
we found the best fit by using $R_{\mathrm{in,qui}}=0.5$\,AU, 
which is the radius of both the dust disk and envelope. 
The variation of the inner radii of the disk and the envelope indicates that a 
dynamical process worked immediately after the outburst, which enlarged the inner dust-free hole from 0.5 to
0.7\,AU on a timescale of a few months.
%
%\subsubsection{Mechanism varying the inner radii of the disk and the envelope} 
%
The derived $R_{\mathrm{in}}=0.5$ and 0.7\,AU radii correspond to 1000\,K and 1500\,K in quiescent state and at the peak brightness phase, respectively. 
These temperature values are close to the canonical dust sublimation temperatures ($\sim1300-1600$\,K). 
However, moving the inner radii of the quiescent model inward to the corresponding sublimation radii, the model underestimates the 
optical data points due to the increased extinction. 
Although similar deviations of the radii were found at other eruptive young systems (e.g., Sipos et al.~\cite{sipos},
Juh\'asz et al.~\cite{juhasz11}, 
Eisner et al.~\cite{eisner11}), the cause of this larger inner radius is not clear.
Varying parameters of the models are shown in Table~\ref{tab:epochs}.

\subsection{\textbf{2005 September: The puzzling MIDI data}}
\label{subsubsec:puzzle}
\label{rt:sept}
\label{rt:rim}
\label{rt:envelope}

For the 2005 March and September epochs the MIDI visibilities give additional constraints for the system
parameters. We attempted to fit the 2005 September SED and visibility values using the
same method followed above. We kept the inner radii unchanged ($R_{\mathrm{in}}=0.7$\,AU ) and 
changed the accretion rate between the two MIDI epochs from  $3.5\times
10^{-6}$\,M$_\odot$/yr to $2.0\times 10^{-6}$\,M$_\odot$/yr. 
Due to the fading of the central source the mid-infrared emitting region is smaller 
and results in slightly increased visibilities. This is just the opposite of what 
the measurements show (shown with open triangles in Fig.~\ref{fig:sedvis}, lower middle panel). 
Clearly, although for the SEDs our first strategy of changing only the
accretion rate and the inner radii of the dust disk and the envelope works, it cannot account for the  changes
in the visibilities. 

In the framework of our static radiative transfer models 
the geometry of the circumstellar environment also had to undergo changes 
in the inner regions of the disk-envelope system 
to explain the lowering of the visibility points.
The change has to take place in the inner regions of the system, because 
the dynamical timescales for changing the structure of the environment farther out is much longer 
than the half-year elapsed between the two MIDI observations.  

What that might help to explain the changes in the visibility data is to assume that
due to the increased heat - because of the enhanced accretion - the scale height of the regions close to the inner edge of the
disk increases in the first phase of the outburst. By introducing such a region to the system
and varying the scale height of the inner part of the disk, i.e., a \textit{variable puffed-up inner rim}, 
we also see variations in the
visibilities. The visibilities are higher if the scale height increases. 
A higher inner rim means an additional hot component in the system, which increases the visibilities. However, this rim also casts more shadow on the disk, 
thus certain regions behind the rim will not be illuminated, which might also affect the visibilities.
We tested several geometries
of the disk to see how strong the effect is and found that this process could only account
for approximately one quarter of the observed decrease, i.e., 0.05-0.08 in the visibilities. 
Therefore we do not discuss these models further. 

Another straightforward idea is to increase  the inner radii of the dusty components from March to September, 
which we expect to have the effect of increasing the mid-infrared emitting  region, i.e., making the source more
resolved.
The inner radii of the envelope and the disk were changed previously and seem to be the parameters that could vary the
easiest (compared to disk scale height or density profile exponents). However, any transportation mechanism that removes
material is much more effective on the significantly less dense envelope,
and would move its inner radius much farther out. Therefore we do not consider any change in the disk in the following. 

In our model that fits the visibilities, the dusty envelope starts only at 3.0\,AU, and the matter
removed from the inner regions is piled up in a thin shell close to the inner edge of the envelope.
This distance corresponds to $\approx600$\,K temperature.
By moving the inner edge outwards, the infrared flux increased slightly, so we had to adjust the accretion rate again, i.e., lowered to 1.6 $\mathrm{M}_{\odot}\mathrm yr^{-1}\times10^{-6}$.
Moving matter outwards, and creating a shell of matter has the additional effect of reducing the extinction, thus increasing the optical-near infrared flux. 
As the results show in Fig.~\ref{fig:sedvis}, improving the fit to the
visibilities deteriorated the SED-fit.
The extinction in this model dropped to $A_\mathrm{V}=$11.5\,mag, which would have led to an unobserved brightening of V1647\,Ori (cf. Fig.~2).  
Although the accretion luminosity decreased in this period, it could hardly compensate for the effect of  
$\Delta A_\mathrm{V}$. 
The uncertainty of the optical part of our model for 2005 September  
prevents us from analyzing of the evolution of the extinction at this epoch. 
We attempt to find the mechanism that can produce the dramatic variation of the envelope radius between  
March and September 2005 in Sect.~\ref{disc:midi1}.

%%%%%%%%%%%%%%%%%%%%%%%%
\section{Discussion}

\subsection{\textbf{The varying accretion rate}}
\label{disc:accr}

The peak accretion rate value we found is on the same order of magnitude 
as the $5-10\times10^{-6}$\,M$_{\odot}/yr$ values derived from  
Br$\gamma$ measurements (AP07) and modeling (Muzerolle et al.~\cite{muze05}). 
During the plateau phase of the outburst, as the object slowly faded, our model accretion rate slightly decreased.  
Both observations and modeling indicate that $\approx$ $10^{-5}$ M$_{\odot}$ material 
was accreted onto the star during the outburst.  

The derived accretion rates fit in the canonical picture of eruptive low-mass YSOs. 
In quiescence the accretion rate of our model ($3.0\times10^{-7}$\,M$_{\odot}/yr$) is similar to those of 
class~I YSOs ($10^{-9}-10^{-7}$\,M$_{\odot}/yr$, White \& Hillenbrand~\cite{white}) or 
class~II objects ($10^{-9}-10^{-6}$\,M$_{\odot}/yr$, Gullbring et al.~\cite{gullbring}, Hartmann et al.~\cite{hartmann98}).  
During the outburst it is $3.0-7.0\times10^{-6}$\,M$_{\odot}/yr$, 
higher than that of quiescent  low-mass YSOs, comparable to that of
FUors ($10^{-6}-10^{-4}$\,M$_{\odot}/yr$, HK96). Intermediate values are expected for EXors (HK96).
Juh\'asz et al.~(\cite{juhasz11}) derived $2.2\times10^{-7}$\,M$_{\odot}/yr$ for the prototype EXor, EX~Lupi, at its outburst peak in 2008.

In the quiescent model 
we derived a similar accretion rate value as AP07 and Muzerolle et al.~(\cite{muze05}) found, 
i.e., a few times $10^{-7}$\,M$_{\odot}/yr$.
Aspin et al.~(\cite{aspin08}) claimed, based on the strength of the Br$\gamma$ emission line,  
that the accretion rate was somewhat higher ($\approx 10^{-6}$\,M$_{\odot}/yr$) after the outburst.  
However, the estimates of the accretion rate derived from the Br$\gamma$ 
measurements strongly depend on the applied value of $A_\mathrm{V}$. Aspin et al.~(\cite{aspin08}) 
considered higher visual extinction than, e.g., AP07. 

As a final refinement to fit the NIR color variations,
%at the optical-NIR wavelengths, 
we also had to
modify the parameters describing the accretion characteristics. At the
outburst peak, the hot spot covers a large part ($\approx$50\%) of the
stellar surface. In quiescence, the hot spot is much smaller, i.e.,
$\lesssim$1\% surface fraction. 
Similar results were determined for other eruptive YSOs (EX~Lupi, Juh\'asz et al.~\cite{juhasz11}, and V1118\,Ori, Audard et al.~\cite{audard}).
Calvet \& Gullbring~(\cite{calvet98}) found fractions of $20\%$ in strongly accreting systems, 
and typically $1\%$ for T~Tau systems.
%These parts of the surface are the sources of a given fraction of the accretion luminosity. 
These surface fractions correspond to 6500 and 15000\,K hot-spot temperatures at the outburst peak and
quiescence, respectively.

\subsection{\textbf{Dynamics of the inner disk and envelope}}
\label{disc:rin}

During the outburst the temperatures were the
highest at the peak, which defines the radius up to which dust had to evaporate. 
The variation of the inner radii of the disk and the envelope suggests
that the physical mechanism enlarging the inner dust-free hole from 0.5 to
0.7\,AU is the evaporation of dust by the increased heating from the
central source.  The intensified wind from the center of the system might
have also played a role in these changes (Reipurth \& Aspin~\cite{ra04}).

\subsection{\textbf{Burst of luminosity or drop of extinction?}}

Owing to the a structural change described previously, the extinction of the system should also change. 
The increased accretion luminosity and the varying extinction determine the total brightening of the object. 
When the heat because of the increased accretion
luminosity evaporates dust particles and creates a larger inner hole, the line-of-sight
extinction decreases, which leads to an additional apparent brightening of the object. 
Since the density in the envelope decreases with radius, most of the extinction comes from
the innermost parts. Therefore, dust evaporation at the inner edge of the envelope can significantly change the total extinction toward the central star.
Similarly, the gradual fading of the central source may allow re-condensation of dust grains in
regions where evaporation took place previously, resulting in an extra dimming.
%\footnote{However, dust re-condensation requires specific conditions (e.g., Gail et al. 
% 2009).}. 
Kun et al.~(\cite{kun}) claimed that such re-condensation was observed at another YSO, PV~Cep .

For the understanding of the outburst,
it is essential to separate to which extent those two agents contributed to the brightening of
V1647\,Ori throughout the eruption.
Some authors (e.g., McGehee et al.~\cite{sdss}, Reipurth \& Aspin~\cite{ra04}) 
concluded that only part of the brightness variation of V1647\,Ori was due to extinction variation, 
e.g., dust-clearing. 
From our models, we derived a change of the extinction 
$\Delta A_\mathrm{V}\approx 4.5$\,mag or $\Delta A_{\mathrm{I}_\mathrm{C}}\approx 2.5$\,mag. 
Compared to the largest measured variation in $\mathrm{I}_\mathrm{C}$ (6.0\,mag, AP07 and Aspin \& Reipurth~\cite{ar09}), 
it leaves more than 3\,mag intrinsic brightening in $\mathrm{I}_\mathrm{C}$.
There are a number of different attempts in the literature to derive the value of extinction and its variation during the outburst. 
Our model result for the quiescent phase ($A_\mathrm{V}\approx$23.4\,mag) agrees quite well with the value derived by Aspin et al. 
(\cite{aspin08}, $A_\mathrm{V}\approx19$\,mag).
The object moved along the reddening path in the NIR color-color diagram during the outburst.  
The excursion corresponds to $\Delta A_\mathrm{V}\approx5$\,mag  (e.g., Reipurth \& Aspin~\cite{ra04}), which is close to our
results. 
However, with different methods, Aspin et al.~(\cite{aspin08}) derived $\Delta A_\mathrm{V}\approx10$\,mag. 
Since the optical extinction values are poorly defined over the outburst, we do not aim at fitting them with 
our model, only attempt a qualitative check of this parameter.  
Our modeling indicated that the extinction is higher in quiescence than at peak brightness.

\begin{table}[h!]
\caption{Varied model parameters for different epochs.}
\centering
\begin{tabular}{lccccc}
\hline\hline
Parameters & \multicolumn{2}{c}{2004} & \multicolumn{2}{c}{2005} & 2003/06\\
		& Mar & Oct 		& Mar & Sept &       quiescent\\
\hline
$\dot{M}$ ($\mathrm{M}_{\odot}\mathrm yr^{-1}\times10^{-6}$)& $7.0$ & $5.5$& $3.5$ & $1.6$ & $0.3 $\\
$R_{\mathrm{in,disk}}$ (AU)&0.7 &  0.7 & 0.7 & 0.7 & 0.5 \\
$R_{\mathrm{in,env}} $ (AU)&0.7 &  0.7 & 0.7 & 3.0 & 0.5 \\
$A_\mathrm{V}$ (mag)      & 18.9& 18.9 & 18.9& 11.5   & 23.4 \\
\hline
\end{tabular}
\label{tab:epochs}
\end{table}

\subsection{\textbf{Scenarios for variations seen by MIDI}}
\label{disc:midi1}
\label{rt:halo}
\label{non-equ}

\paragraph{A blown-up spherical cavity} The expected structural changes can either occur relatively slowly during the March-September interval
or in a few weeks' time (cf. the light curves in Fig.~\ref{img:lightcurves}). 
Even in the latter case such a change of the envelope seems to be plausible (corresponding to $100-200$\,km/s velocities).
During most of this period the object was slowly fading, therefore we would expect the decrease of the inner  
(dust evaporation) radii both of the disk and the envelope rather than any increase. 
Combet \& Ferreira~(\cite{combet}) showed that at the accretion rate of $10^{-6}$\,M$_\odot$/yr, a strong disk wind, which can be the source of an outflow,
is launched from the inner disk up to 2-3\,AU. Thus the disk wind might be responsible for the clearing of the inner envelope. 
However, during the 2005 March-September period the wind is expected to become weaker since  
its strength should be proportional to the accretion activity.  
Furthermore, the wind that was present from the beginning of the outburst  
and had no such effect before might only have cleared out the cone in the reflection 
nebula when the outburst started (Reipurth \& Aspin~\cite{ra04}). 
Therefore the stellar or disk wind is unlikely to have produced this clearing.  
 
Another possibility is that some temporary outflow changed the inner structure of the system and moved the inner edge of the envelope outward. 
Spectroscopic observations from early 2006 showed evidence of a short outflow activity (Brittain et al.~\cite{brittain}).  
Furthermore, detection of forbidden optical lines indicated the presence of shocked gas in early 2006 (Fedele et al.~\cite{fedele_sp}).  
These lines can be tracers of a Herbig-Haro object  
(HHO)\footnote{Although HHOs are thought to be related 
to mass accretion, they are not typical in the environment of eruptive YSOs (HK96). 
Note that Eisl\"offel \& Mundt~(\cite{eis}) identified V1647\,Ori as the 
driving source of HH23 -- presumably ejected some thousands of years ago --   
which is located 155\arcsec north of the young star and is close to axis of the  
nebula.  }.  
If such an event were responsible for the change of the inner radius of the envelope ($R_{\mathrm{in,env}}$), based on the beforementioned findings, it should be 
connected to the final decrease of the accretion rate. 
However, Brittain et al.~(\cite{brittain}) proposed that because the inner radius of the accretion disk decreases together with the decreasing accretion rate,  
the re-structuring of the inner disk that produced the (CO) outflow might have occured later.  

During an outflow activity or the appearance of an HHO, the disk structure can be very different than that of a standard accretion disk.   
There is a significant difference in all parameters  
between those of a standard accretion disk and a jet-emitting disk (Combet \& Ferreira~\cite{combet}). 
Reproducing such events are beyond the limits of our static modeling and can be a source of discrepancy.  
 
By 2006 February, the object returned to its quiescent brightness at all wavelengths.
This suggests that the structural changes in the circumstellar environment of V1647\,Ori  
should have been reversible on the timescale of some months. 
It means that the material should fill the evacuated spherical cavity on this timescale via accretion,
which would require an inward radial velocity of 10-20 km/s.
%It corresponds to a velocity of 10-20 km/s. 
In our modeling it means that the inner radii of both the disk and the envelope should move back close to the quiescent 
values. 
Matter of the disk and the envelope might have moved inward due to the accretion and thus could fill the cleared regions.  
However, the dynamical timescale at $\sim$3\,AU is about seven years, much longer than half a year. 
Therefore dust re-condensation might also have played a role here.

\paragraph{The disappearance of a warm halo}
In the next scenario we again consider a large spherical inner cavity in the envelope as in
Sect.\,\ref{rt:envelope} ($R_{\mathrm{in,env}}=3.0\,\mathrm{AU}$ while 
$R_{\mathrm{in,disk}}=0.7\,\mathrm{AU}$), but
we assume that the cavity has been produced at the early phases of the outburst for instance by stellar
or disk winds (see e.g., Clarke et al.~\cite{clarke05}).  If dust was
lifted from the disk above it by wind (Vinkovi\'c \& Jurki\'c~\cite{vinkovic}, Sitko et al.~\cite{sitko}), the dust could fill
the inner cavity of the envelope. Note that such a halo cannot be distinguished from the inner envelope in our models, but represent a different mechanism. The warm dust halo present in the system during the plateau phase makes V1647\,Ori look more compact at the first MIDI epoch. The apparent
change of the visibility data could then be caused by the waning wind. As the accretion process and thus the
wind becomes weaker, less and less dust is fed into the cavity, while the dust moved there previously is blown
farther outward. In this case most of the dust - which made the object compact in 2005 March - is cooled
below the temperatures corresponding to the mid-infrared wavelengths.  By 2005 September the cavity is
practically cleared. Without carrying out new modeling, we can consider the results of the radiative transfer
modeling described above. This scenario could explain the transition between 2005 March and September seen in the
data  and also the short timescale variability of the mid-infrared emission.

\paragraph{An out-of-equilibrium system} Finally, we further speculate to find other alternatives for our series of static models.
Between 2005 March and September
the heated disk area should shrink gradually because of the fading of the central illuminating source.  
In contrast, the visibilities of V1647\,Ori apparently did not follow such a scenario.  
The above hypothesis would predict a smaller -- thus less resolved -- source at our second MIDI epoch.   
Our measurements, however, revealed the opposite behavior: the visibilities decreased 
between the first and the second epochs. Since the shapes of the two visibility curves 
are similar (Fig. 1.), and their ratio is almost independent of the wavelength ($\approx$1.4--1.5), 
a simple qualitative picture might explain the data. 
One could assume that the 
system consists of two components: one compact central and one extended component (e.g., approximated by a Gaussian brightness distribution as in Sect. 3.1). If the resolved component did not change  
between the two MIDI epochs, but the emission of the central unresolved source  
had significantly dropped in the same period, the emission of the system would have become less peaked,  
i.e., relatively more resolved. 
This might be a simplistic picture of what caused the decrease of the measured visibilities.

The constancy of the extended emission component could easily be explained 
if the $\sim10\,\mu$m flux arose from optically thick regions, 
whose temperature is not adjusted rapidly to the changing central illumination field. 
Although we see variations of the mid-IR brightness of the system on weekly timescales, 
that of an optically thick component could not change significantly because the corresponding timescales (Chiang \& Goldreich~\cite{cg97}) 
exceed even the difference between the epochs of the MIDI observations.  
However, according to Muzerolle et al.~(\cite{muze05}), the contribution of an optically  
thick accretion disk to the mid-infrared emission of V1647\,Ori in outburst  
is small, and the $\sim$10\,$\mu$m flux mainly arises from 
an envelope. In this envelope the temperature 
of the dust grains must be quickly adjusted to the external radiation field.  
This suggestion is supported by the fact that the 
mid-infrared flux of V1647\,Ori had increased remarkably in less than a few months,  
between the beginning of the eruption and the Spitzer measurements in March 2004. 
Thus the invariability of the extended emitting component in the simple picture above is not  
straightforward to explain.  

A qualitative argument leads to a possible solution in which the decrease seen in the MIDI visibilities 
is caused by the fading of the central source, which is not immediately followed by the fading of the outer regions. However, this effect should be confirmed by dynamical modeling and is beyond the scope of the present paper.

\section{Conclusions}

We performed interferometric and photometric observations of V1647\,Ori during its 2003-2006 outburst to investigate 
the temporal evolution of its circumstellar structure and physical processes related to the eruption.
In addition to the general fading of the object - shown by the multi-wavelength photometric data and 
archival mid-infrared spectroscopy, - short timescale variations were also observed.
Optical-infrared SEDs at five epochs were compiled. 
Our radiative transfer modeling, with a smoothly decreasing accretion rate as a major varying parameter, 
provided good fits of the SEDs at different stages of the outburst. It is important to note that the inner radii of the dust disk and envelope also
had to increase during the transition from quiescence to the outburst peak.
The latter finding is clear evidence of dynamical variations in the inner circumstellar
environment of V1647\,Ori. This dust clearing is likely caused by the evaporation of the dust grains 
due to the outburst heat.

High angular resolution spatial information were also considered in the model-fitting procedure. 
VLTI/MIDI data obtained at two epochs, during both the slow and rapid fading stages, show a considerable change of the 
circumstellar structure. In constrast to our expectations, based on our model sequence, the object looked more resolved
at the second epoch. 
One possible explanation can be a rapid removal of dust from the inner 3\,AU of the envelope, possibly caused by wind or outflow processes. In one case this spherical cavity is produced at the end of the outburst phase (the blown-up cavity scenario), or alternatively at the beginning (our warm dust halo scenario). 
Finally, we may also see V1647\,Ori in a non-equilibrium situation in 2005 September, 
when the sudden fading of the central source was not yet followed by the fading of the optically thick circumstellar material.

In general, our modeling showed that the circumstellar environment of V1647\,Ori can be described by a disk and envelope system with parameters that are typical for embedded low-mass YSOs. This finding supports the hypothesis that eruptive YSOs are not peculiar objects, but represent an important phase in the evolution of all low-mass YSOs.

\begin{acknowledgements} 
The authors thank the anonymous referee for his/her comments, which improved the manuscript.
The authors thank the ESO/VLTI staff for executing the observations in service mode, 
Timea Csengeri for obtaining photometry at IAC,  
Aurora Sicilia-Aguilar and Kees Dullemond for useful discussion,  
Karl Gordon for the method of the MIPS data reduction.  
LM is grateful to Walter Jaffe and Rainer K\"ohler for helping with the MIDI data reduction.  
The research leading to these results has received funding from the
European Community's Seventh Framework Programme under Grant Agreement
226604.
LM and NS are thankful for the support of the Fizeau Exchange Visitor Program through the European Interferometry Initiative 
(EII) and OPTICON (an EU funded framework program, contract number RII3-CT-2004-001566).  
The research of \'A.K. is supported by the Netherlands Organization for Scientific Research.
Financial support from the Hungarian OTKA grants K81966, K101393 and NN102014 are acknowledged.
\end{acknowledgements}

%%%%%%%%%%%%%%% 
%%%%  Appendix 
%%%%%%%%%%%%%%% 
 
%\Online 
 
\begin{appendix} %online appendix 
%\appendix 
\section{Data reduction with different versions of MIA+EWS} 
\label{app:dr} 
 
%1.2 and 1.3 vs 1.5 and 1.6 

Since we aim at investigating temporal changes of the source structure, 
and the first MIDI datasets were reduced with the 1.3 version of MIA+EWS, 
we repeated the data reduction with the 1.6 version.  It turned 
out that longward of $11\mu$m the calibrated visibilities differ 
significantly with respect to the previous reduction, 
i.e., are higher than were shown in Paper~I. At shorter 
wavelengths the results are consistent. Nonetheless, the results with EWS and 
MIA agreed well in Paper~I. The EWS and MIA results are also consistent for the second epoch data 
(Sect.~\ref{dr:midi}). As another test, we also reduced the 
second MIDI dataset with the older versions of the software. The results 
show a similar bias. In the case of MIA, the difference comes mainly from 
two changes mage since the first release.
First, newer MIA versions use the photometry routine of EWS instead of 
its own (used in earlier versions). Second, these new versions of MIA 
fit fewer parameters of the mask than the older MIA version did. 
In the case of EWS, the 
difference is that its newer versions handle the data 
where the OPD difference is close to zero, differently from older EWS versions, i.e., these data are not considered anymore. Since all these changes of the softwares were 
made to improve the data reduction, we present calibrated 
visibility curves for both epochs produced wih MIA (1.6)  and considered 
these for the radiative transfer modeling throughout the paper.  

\listofobjects

\end{appendix} %Last line of the first/second online appendix 

%%%%%%%%%%%%%%%%%%%%%%%%%%%%%%%%%%%%%%%%%%%%%%%%%%%%%%%%%%%%%%%%%%%%%%%%%%%% 
 
%\end{document} 

%%%%%%%%%%%%%%%%%%%%%%%%%%%%%%%%%%%%%%%%%%%%%%%%%%%%%%%%%%%%%%%%%%%%%%%%%%%%%%%%%%%%%
 
\end{document}